\documentclass[traditabstract,a4paper]{aa}
\usepackage{amsmath}
\usepackage{graphicx}
\usepackage{color}
\usepackage{natbib}

\bibpunct{(}{)}{;}{a}{}{,}

\begin{document}

\nocite{*}

\author{Edo van Uitert \inst{\ref{inst0},\ref{inst1},\ref{inst2}}\thanks{
        E-mail: vuitert@ucl.ac.uk} \and David G. Gilbank \inst{\ref{inst3}}  \and Henk Hoekstra \inst{\ref{inst1}} \and Elisabetta Semboloni \inst{\ref{inst1}} \and Michael D. Gladders \inst{\ref{inst4}} \and H.K.C. Yee \inst{\ref{inst5}}}
\institute{Argelander-Institut f\"ur Astronomie, Auf dem H\"ugel 71, 53121 Bonn, Germany \label{inst0} \and Leiden Observatory, Leiden University, Niels Bohrweg 2, NL-2333 CA Leiden, The Netherlands \label{inst1} \and University College London, Gower Street, London WC1E 6BT, UK \label{inst2} \and  South African Astronomical Observatory, PO Box 9, Observatory, 7935, South Africa \label{inst3} \and The Department of Astronomy and Astrophysics, and the Kavli Institute for Cosmological Physics, The University of Chicago, 5640 South Ellis Avenue, Chicago, IL 60637, USA \label{inst4}  \and Department of Astronomy and Astrophysics, University of Toronto, 50 St. George Street, Toronto, Ontario, M5S 3H4, Canada \label{inst5}}

\title{Weak-lensing-inferred scaling relations of galaxy clusters in the RCS2: mass-richness, mass-concentration, mass-bias, and more}

\titlerunning{Weak-lensing-inferred scaling relations of galaxy clusters in the RCS2}

\abstract{ We study a sample of $\sim$10$^4$ galaxy clusters in the redshift range 0.2$<$$z$$<$0.8 with masses $M_{200}$$>$$5\times10^{13} h_{70}^{-1} M_{\odot}$, discovered in the second Red-sequence Cluster Survey (RCS2). The depth and excellent image quality of the RCS2 enabled us to detect the cluster-mass cross-correlation up to $z$$\sim$0.7. To obtain cluster masses, concentrations, and halo biases, we fit a cluster halo model simultaneously to the lensing signal and to the projected density profile of red-sequence cluster members, because the latter provides tight constraints on the cluster miscentring distribution. We parametrised the mass-richness relation as $M_{200}=A\times(N_{200}/20)^{\alpha}$ and find \mbox{$A=(15.0\pm0.8) \times10^{13} h_{70}^{-1} M_{\odot}$} and $\alpha=0.73\pm0.07$ at low redshift (0.2$<$$z$$<$0.35). At intermediate redshift (0.35$<$$z$$<$0.55), we find a higher normalisation, which points towards a fractional increase in the richness towards lower redshift caused by the build-up of the red sequence. The miscentring distribution is well constrained. Only $\sim$30\% of our BCGs coincide with the peak of the dark matter distribution. The distribution of the remaining BCGs are modelled with a 2D-Gaussian, whose width increases from 0.2 to 0.4 $h_{70}^{-1}$Mpc towards higher masses. The ratio of width and $r_{200}$ is constant with mass and has an average value of $0.44\pm0.01$. The mass-concentration and mass-bias relations agree fairly well with literature results at low redshift, but have a higher normalisation at higher redshifts, possibly because of selection and projection effects. The concentration of the satellite distribution decreases with mass and is correlated to the concentration of the halo.
}


\maketitle


\section{Introduction}  
\hspace{4mm} Observations of galaxy clusters provide a wealth of astrophysical and cosmological information. A key quantity of clusters is the mass, because it determines the relative importance of various processes such as AGN feedback. Furthermore, given a large sample of cluster masses, the cluster mass function can be determined and compared to simulations in order to constrain cosmological parameters, such as the normalisation of the matter power spectrum, $\sigma_8$, and the cosmological matter density, $\Omega_{\rm M}$ \citep[e.g.][]{Evrard89,White93}; if the redshift baseline of the sample is sufficiently large, the dark energy equation of state can be constrained \citep[e.g.][]{Voit05,Allen11}.\\
\indent The mass of a cluster is not a direct observable, but can be determined with a variety of techniques. The velocity distribution of cluster members have been used to derive dynamical mass estimates \citep[e.g.][]{Vandermarel00,Lokas06}, but these observations are generally expensive since they require spectroscopic observations of many cluster members. Additionally, assumptions on the satellite orbits are needed to convert the velocity dispersions into a mass estimate. X-rays observables can also be used to estimate the mass \citep[see][for a review]{Ettori13}, under the assumption that the hot cluster gas is in hydrostatical equilibrium. The results of \citet{Mahdavi08,Mahdavi13} support the results from hydrodynamical simulations \citep[e.g.][]{Nelson14} that clusters are generally not in hydrostatical equilibrium, which biases the \mbox{X-ray-based} mass estimates. Another powerful method of obtaining cluster masses is weak gravitational lensing. \\
\indent In weak lensing, the distortion of the images of faint background galaxies (sources) due to the gravitational potentials of intervening structures (lenses) is measured. This signal is proportional to the excess surface mass density, which can be modelled to obtain the mass. Weak lensing does not rely on direct tracers of the potential; the distortion can be measured for any lens, out to large radii where no visible tracers can be used. Additionally, the weak lensing signal does not depend on the physical state of the matter in the clusters, and no assumptions have to be made (e.g. virial equilibrium) to measure the total projected mass. Weak lensing has been used to determine the mass of individual massive low-redshift clusters \citep[e.g.][]{Hoekstra07,Okabe10,Hoekstra12,Applegate14,Gruen13,Umetsu14,Kettula15,Hoekstra15}, as well as the average mass of samples of clusters and galaxy groups by stacking their lensing signals \citep[e.g.][]{Mandelbaum_clus06,Sheldon09I,Covone14,Ford15}. \\
\indent From the weak-lensing signal, many cluster properties can be extracted, such as the cluster mass, concentration, halo bias, and miscentring distribution. The relation between these parameters can help constrain models of cluster physics. To determine the mass function, however, we need mass estimates of a large number of clusters. The lensing signal of all but the most massive clusters is generally noisy; only by stacking the signal of samples of clusters can the average mass be robustly constrained. A common solution is to determine how an observable cluster property scales with mass and can serve as a mass proxy. The Sunyaev-Zeldovich effect has been used \citep[e.g.][]{Williamson11} and appears particularly useful for estimating the masses of massive clusters at high redshifts through scaling relations. Another observable property is the richness, which has the advantage that it can be determined from the same multi-colour imaging data that is used for the lensing analysis. \\
\indent To determine the richness of a cluster, it is necessary to distinguish cluster galaxies from fore- and background galaxies and, if necessary, correct for contamination. Cluster members can be identified if their redshift or velocity dispersions are available, which requires either spectroscopy or observations in many bands for reliable photometric redshifts. Alternatively, cluster members can be identified using their colours as the majority of early-type galaxies in a cluster populate a narrow range in colour-magnitude space, that is, the E/S0 ridge line or the red sequence \citep{Gladders00}. The advantage of the latter is that observations in only two bands suffice, which makes it cheap and particularly suited for the automated detection of clusters in large imaging surveys \citep[e.g.][]{Gladders05}. Additionally, only a few field galaxies reside in this regime of colour-magnitude space, which reduces the contamination. \\ 
\indent In this paper we use optical imaging data from the second Red-sequence Cluster Survey \citep[RCS2; ][]{Gilbank11}, both to detect clusters and determine their richness, as well as to measure their lensing signals. The survey design was chosen so as to optimise the detection of a large number of clusters using a red-sequence method \citep{Gladders00}. In total, $\sim$$10^4$ clusters have been detected in the RCS2, spread over a wide range in optical richness, with redshifts 0.2$<$$z$$<$0.9. In contrast, the redshift range of the maxBCG cluster sample \citep{Koester07}, a catalogue of 13\hspace{0.5mm}823 clusters that has been detected in the Sloan Digital Sky Survey \citep[SDSS;][]{York00}, covers a redshift range of 0.1$<$$z$$<$0.3, which limits its use for evolutionary studies. The redshift range of clusters in the RCS2, combined with the excellent lensing quality of the data, makes the RCS2 well suited to this purpose. \\
\indent The outline is as follows. In Sect. \ref{sec_analysis}, we present the various steps of the analysis: we discuss the cluster detection and richness estimates (\ref{sec_clussel}), provide details of the lensing measurement (\ref{sec_shape}), and discuss our halo model (\ref{sec_hm}). We highlight our novel approach of including the cluster-satellite correlation to constrain the miscentring distribution in Sect. \ref{sec_crs}. We present the mass-richness relation in Sect. \ref{sec_mr}, the cluster miscentring distribution in Sect. \ref{sec_msc}, the mass-concentration relation in Sect. \ref{sec_mc}, the satellite distribution in Sect. \ref{sec_msat} and the mass-halo bias relation in Sect. \ref{sec_mb}. We conclude in Sect. \ref{sec_concl}. Throughout the paper we assume a WMAP7 cosmology \citep{Komatsu11} with $\sigma_8=0.8$, $\Omega_{\Lambda}=0.73$, $\Omega_{\rm M}=0.27$, $\Omega_{\rm b}=0.046$ and $h=0.7$ the dimensionless Hubble parameter. All distances quoted are in physical (rather than comoving) units unless explicitly stated otherwise.


\section{Analysis \label{sec_analysis}}

\hspace{4mm} The RCS2 is a nearly 900 square degree imaging survey in three bands ({\it g$'$, r$'$} and {\it z$'$}) carried out with the Canada-France-Hawaii Telescope (CFHT) using the one square degree field of view camera MegaCam. The primary imaging data covers 740 square degrees of sky, divided in 13 patches. The survey area can be split into 145 blocks of contiguous non-overlapping 2$\times$2 degrees of sky, which we use to estimate the bootstrap covariance matrices of our measurements as discussed later on. Hence in total we use 580 square degrees. The lensing analysis is performed on the eight minute exposures in the {\it r$'$}-band ({\it $r'_{\rm{lim}}$$\sim$}24.3), which is best suited for lensing with a median seeing of $0.71''$. \\
\indent The photometric calibration of the RCS2 is described in detail in \citet{Gilbank11}, the lensing analysis in \citet{VanUitert11}. For details, we refer the reader to these works. In short, we measured the shapes of 2.2$\times 10^7$ galaxies using the KSB method \citep{Kaiser95,LuppinoK97,Hoekstra98,Hoekstra00}, which corresponds to a source number density of 6.3 arcmin$^{-2}$.  Two major improvements to the lensing analysis were introduced in \citet{Cacciato14,VanUitert15}: we used the photometric redshift catalogues from \citet{Ilbert13} instead of the catalogues from \citet{Ilbert09} to estimate the source redshift distribution; and secondly, we introduced a correction scheme to account for a multiplicative bias in our KSB method (due to noise bias \citep[][]{Kacprzak12,Melchior12,Refregier12}, and galaxy blends), which affects the lensing measurement \citep{Hoekstra15}. \\


\subsection{Cluster detection \label{sec_clussel}}
\begin{figure}
  \resizebox{\hsize}{!}{\includegraphics[angle=270]{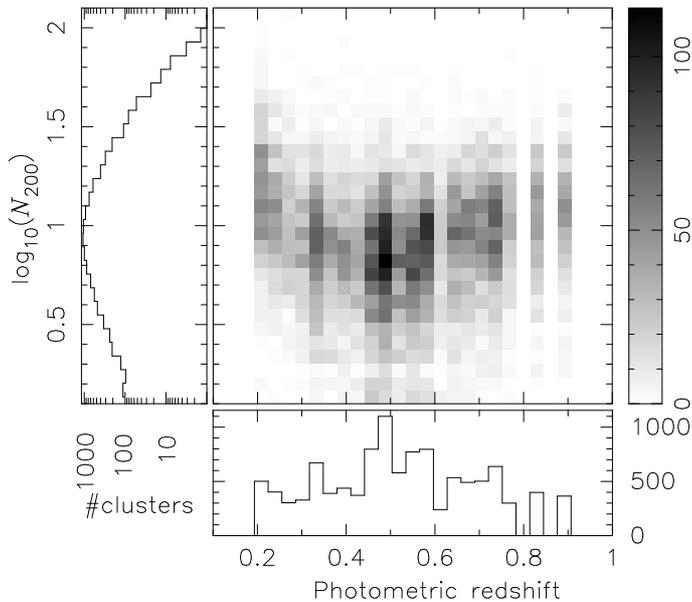}}
  \caption{Redshift versus $\log_{10}(N_{200})$, the logarithm of the number of early-type cluster members brighter than $M^\star+1$ inside $r^{\rm gal}_{200}$. The clusters cover a wide range in richness and redshift, and are therefore very well suited to studying the redshift dependence of the mass-richness relation. }
  \label{plot_z_n200}
\end{figure}
\begin{figure}[t]
  \resizebox{\hsize}{!}{\includegraphics[angle=270]{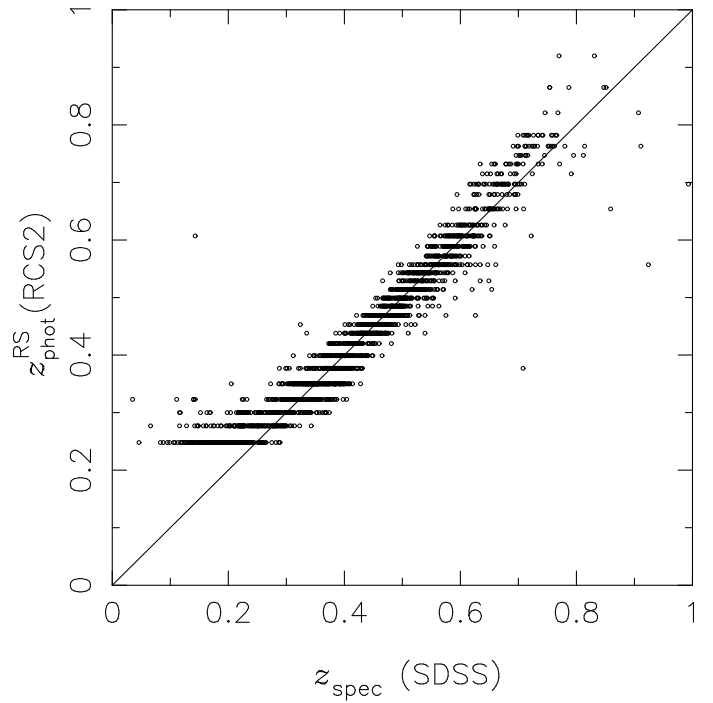}}
  \caption{Comparison of our `red-sequence' redshifts to the spectroscopic redshifts from SDSS. }
  \label{plot_zcomp}
\end{figure}
\begin{figure}[t]
  \resizebox{\hsize}{!}{\includegraphics{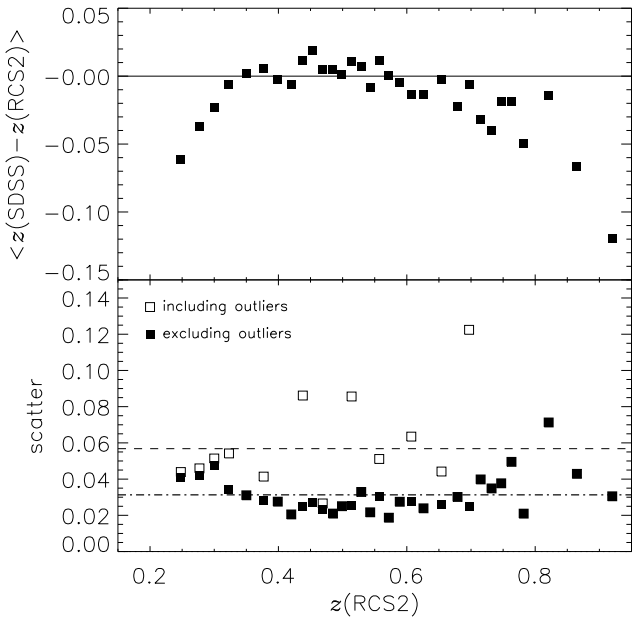}}
  \caption{Mean difference between our `red-sequence' redshifts and the spectroscopic redshifts from SDSS ({\it top}). Scatter between the redshifts after accounting for the bias ({\it bottom}). Open symbols indicate the scatter for all matches, solid ones are obtained after removing the outliers ($|z_{\rm RCS2} - z_{\rm SDSS}|>0.15$). The dashed/dotted-dashed line shows the mean scatter including/excluding outliers. We correct the mean redshift bias in our analysis.}
  \label{plot_zbiasscat}
\end{figure}
\begin{figure}
  \resizebox{\hsize}{!}{\includegraphics[angle=270]{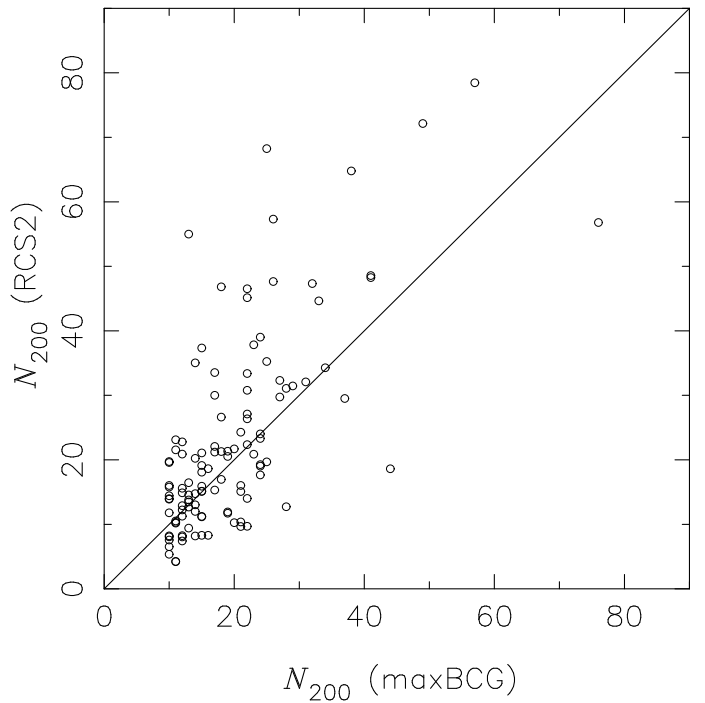}}
  \caption{Comparison of the cluster richnesses of 114 matched clusters from maxBCG and RCS2. The solid line shows the one-to-one relationship. }
  \label{plot_richcomp}
\end{figure}

\indent Galaxies clusters are identified using a modified version of the algorithm presented in \citet{Lu09}. This is a simplified version of the \citet{Gladders05} red-sequence cluster-finding method. The main modifications are described in Gilbank et al. (in prep.), detailing how the method was applied to RCS2 data. Briefly, the significance of a candidate cluster overdensity is determined from a count of (colour-selected) galaxies within a circle with a radius of 0.5 Mpc. Unlike the \citet{Gladders05} method, no magnitude weighting is applied to the galaxies, and the uncertainties in the cluster and field counts are estimated directly from Poisson statistics. Colour slices are built at regular colour intervals (which leads to irregular bins in photometric redshift), and each slice overlaps its neighbour by a quarter of the slice width. For RCS2, the method is identical except that the model colour--redshift relation was transformed to the RCS2 filters. One additional modification for RCS2 involves the centring adopted, which aims to locate the BCG via a simple two-step approach, as discussed in Appendix \ref{ap_bcg}.  \\
\indent Richnesses are estimated using an approach similar to the one outlined in \citet{Koester07}, \citet{Hansen05,Hansen09}, and \citet{Johnston07} for the maxBCG cluster sample. Firstly, the number of red-sequence galaxies brighter than M$^\star+1$ within an aperture of radius one Mpc is counted.  M$^\star$ is estimated from simple stellar population synthesis models, as described in \citet{Lu09}. This count is then used to estimate $r_{200}$ (the radius inside of which the density is 200 times the critical density, $\rho_c$) for each cluster using the relation \hbox{$r_{200}^{\rm gal} = 0.156 N_{\rm gal}^{0.6}$ Mpc} from \citet{Hansen05}. The number of red-sequence candidates brighter than M$^\star+1$ within $r_{200}^{\rm gal}$ gives $N_{200}$, our richness estimate. Since we also apply a background subtraction, $N_{200}$ are non-integer values. \\
\indent The maxBCG papers listed above use a variety of slightly different scaling relations which vary in both exponent and normalisation depending on precisely how $r_{200}$ was derived. The relation we choose is similar to that given in \citet{Hansen05}, although their radius is quoted in $h_{100}^{-1}$Mpc and here we do not correct for the different cosmology, given the range of values in the normalisation from method-to-method. In practise, a different normalisation should just lead to counting within a different fraction of $r_{200}$. Therefore, mass--richness estimates within a single cluster sample should still be internally self-consistent, and to compare between samples it should suffice to apply a constant systematic scaling. \\
\indent The distribution of cluster redshifts and richnesses are shown in Fig. \ref{plot_z_n200}. To assess the accuracy of the `red-sequence' redshifts, we match the BCGs to the galaxies with spectroscopic redshifts from the ninth data release of the SDSS \citep[DR9;][]{Ahn12}. In total, we find 2212 matching galaxies. We compare the redshifts in Fig. \ref{plot_zcomp} and find that they agree quite well. Note that the stripes of `red-sequence' redshifts reflect the redshifts slices used to detect the clusters. Only at redshifts below $z$$<$0.3 and above $z$$>$0.7, the `red-sequence' redshifts are slightly overestimated. The average difference in each redshift slice is shown in Fig. \ref{plot_zbiasscat}. At low redshift, the bias is likely the result of only using red-sequence templates up to $z$=0.248; at lower redshifts, the colours become degenerate. The cause of the bias at $z$$>$0.7 is less obvious, but seems to indicate that the initial calibration of the red-sequence templates using spectroscopic data was affected by outliers (either mismatches or matches that sampled regions with poorer than average photometry). We examined these clusters in detail and decided there was no obvious reason to reject them from the analysis. We correct the cluster redshifts for this bias by fitting a third-order polynomial to these matching galaxies, and apply the same shift to all our clusters. The richnesses are recomputed using the corrected redshifts. Note that Fig. \ref{plot_z_n200} shows the corrected redshifts and the corrected richnesses, the quantities used in this work. \\
\indent After correcting the redshift bias, we compute the scatter and show it in the lower panel of the same figure. When we remove obvious outliers using \mbox{$|z_{\rm RCS2} - z_{\rm SDSS}|>0.15$}, which are likely mismatches between the photometric cluster and the galaxy with spectroscopy, the scatter has a value of $\sim$0.03 and does not vary much with redshift. Note that for `red-sequence' redshifts larger than 0.8, the bias and scatter cannot be well determined because of the low number of matches. For this work that is not important, as we restrict our analysis to clusters with $z$$<$0.8. \\
\indent  To evaluate our richness estimates, we match our clusters to the maxBCG catalogue. We find 114 matches and compare the $N_{200}$ estimates in Fig. \ref{plot_richcomp}. The richnesses of clusters in the RCS2 appear systematically larger at the high-richness end. This may partly be attributed to an improved deblending in the RCS2 because of the better observing conditions. The scatter between the estimates is quite large, part of which may be attributed to particular settings in the cluster detection algorithm that deal with mergers and projections on the sky.  \\
\indent Not all detections in the cluster catalogue are real clusters: a fraction of the clusters may actually correspond to a chance projection of galaxies rather than to a real cluster. These false detections have presumably a different lensing mass than the real clusters of that richness, but since both the richness and mass are affected simultaneously, the bias on the scaling relation is expected to be small. The fraction of real clusters is called the purity, which is generally a function of richness and redshift, but also depends on the cluster detection algorithm. Therefore, to determine the actual value of the purity for our cluster sample, we need to apply the detection algorithm to mock data that mimic the RCS2, which has not yet been done. The false positives do not add random noise, but a coherent (but likely lower) lensing signal. How large the impact is on the lensing mass, needs to be addressed with simulations. Note that the detection significance in our cluster finder is high, favouring a high purity over a high completeness. \\
\indent Figure \ref{plot_z_n200} shows that our cluster sample is incomplete at the low richness end. In principle, this should not affect our results, as the average lensing signal only becomes noisier if we miss clusters in our sample. However, to use the cluster sample to constrain cosmological parameters, a detailed knowledge of the completeness function is a prerequisite.


\subsection{Lensing measurement \label{sec_shape}}
\begin{figure}
  \resizebox{\hsize}{!}{\includegraphics{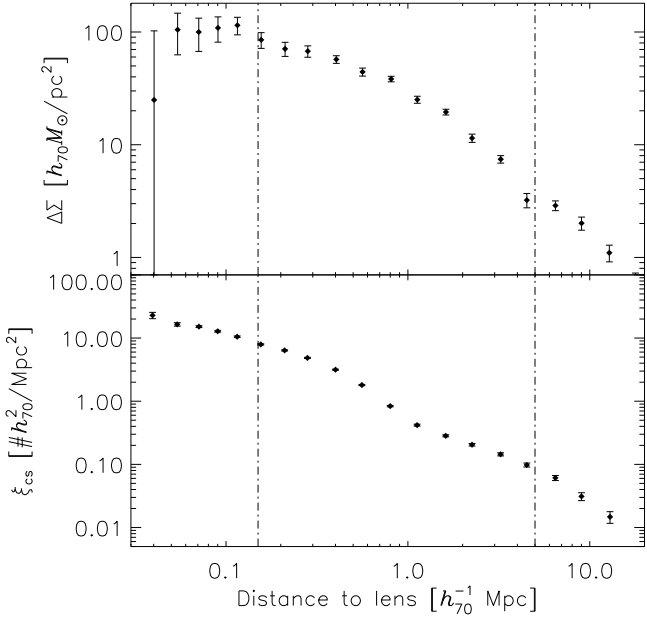}}
  \caption{Stacked lensing signal ({\it top}) and cluster-satellite correlation ({\it bottom}) measured for all clusters with $N_{200}>2$ and $0.2<z<0.8$ in the RCS2. The vertical dot-dashed lines indicate the fitting range for the cluster halo model. Both measurements are shown for illustration only.}
  \label{plot_centdiff}
\end{figure}

\hspace{4mm} The shapes of source galaxies are distorted by the gravitational potentials of clusters. In the regime where the surface mass density is sufficiently small, the gravitational shear can be approximated by averaging the ellipticities of source galaxies \citep[for a review of cluster lensing, see][]{Hoekstra13}. We determine the tangential component of the shear in radial bins centred on the BCGs,
\begin{equation}
  \langle\gamma_t\rangle(r) = \frac{\Delta\Sigma(r)}{\Sigma_{\mathrm{crit}}},
\end{equation}
which is related to the surface density contrast,
\begin{equation}
 \Delta\Sigma(r)=\bar{\Sigma}(<r)-\bar{\Sigma}(r),
 \label{eq_ds}
\end{equation}
 the difference between the mean projected surface density enclosed by $r$ and the mean projected surface density at a radius $r$. The shear signal at small scales around massive clusters is so large that it no longer can be approximated as being linearly related to the galaxy ellipticities. We accounted for this when we computed the models (see Section \ref{sec_hm}). $\Sigma_{\mathrm{crit}}$ is the critical surface density:
\begin{equation}
  \Sigma_{\mathrm{crit}}=\frac{c^2}{4\pi G}\frac{D_{\rm s}}{D_{\rm l}D_{\rm ls}},
\end{equation}
with $D_{\rm l}$, $D_{\rm s}$ and $D_{\rm ls}$ the angular diameter distance to the lens, the source, and between the lens and the source, respectively \citep{Bartelmann01}. Since we lack redshifts for the sources, we select galaxies with 22$<$$m_{r'}$$<$24 that have a reliable shape estimate (ellipticities smaller than one, no SExtractor flag raised) as our source sample. To determine their approximate redshift distribution, we apply identical magnitude cuts to the publicly available photometric redshift catalogue of \citet{Ilbert13}. We use this redshift distribution to compute the mean lensing efficiency, $\langle D_{\rm ls}/D_{\rm s}\rangle$, as a function of lens redshift, accounting for the changing average weight of the sources as a function of apparent magnitude, and accounting for the impact of outliers in the catalogue. This procedure is outlined in Appendix C of \citet{Cacciato14}.\\ 
\indent In Appendix B of \citet{VanUitert15}, we present a test of the robustness of our measurement algorithm. In short, we measure the lensing signal using different source magnitude cuts, and for each cut we recompute the lensing signal (after deriving new lensing efficiencies, random signals, noise bias corrections and source galaxy contamination corrections). Both at low and high redshifts, the resulting lensing measurements are consistent, suggesting that this method of measuring tangential shear is robust.\\ 
\indent Here we present results from an additional test. Since we only select galaxies with a flux radius that is at least 10\% larger than the local PSF size, it is possible that we systematically remove the smallest, highest redshift galaxies from our analysis; we do not account for that when we compute the lensing efficiencies, which could potentially lead to biases. We check that by removing the 10\% smallest objects (in terms of $r_h$) from the photo-z catalogues of \citet{Ilbert13}, and recompute the lensing efficiencies. At low lens redshifts, the results are practically unchanged. At $z=0.7$, the highest mean lens redshift that we study in this work, the lensing efficiencies only decrease by 3\%. Removing the smallest 40\% of objects (much more than we do in practice) leads to a decrease of 8\%. Hence the effect is not completely negligible, but it is smaller than our statistical errors and unlikely to lead to a significant bias. \\
\indent As mentioned before, the cluster redshifts have been corrected for a mean bias. This correction is uncertain for redshifts larger than 0.8 due to the low number of matches. Hence we restrict ourselves to clusters at $z$$<$0.8. The scatter of the lens redshifts also affects the lensing measurement. We estimate the effect in Appendix \ref{ap_zscat}, and find that the impact on the lensing signal is at the per cent level. Therefore, it can be safely ignored.  \\
\indent The distortions induced by weak lensing are much smaller than the intrinsic ellipticities of the sources. The lensing measurement of a single cluster is therefore generally very noisy. To improve the signal-to-noise, the lensing signal is stacked for a sample of clusters that have similar properties (e.g. within a certain richness range). Stacking the lensing signal has the additional advantage that the contribution from uncorrelated structures, as well as from potential small-scale residual systematics, averages out. The lensing signal of the total cluster sample that is used in this work is shown in Fig. \ref{plot_centdiff}. \\
\indent On large scales, residual systematics might affect the lensing signal. These systematics are commonly removed by measuring the lensing signal around random points and subtracting that from the real signal. Such a correction could also remove some real signal, hence we do not apply this correction, but use it as a test of the quality of our catalogues. The mean random signal, averaged over the 145 non-overlapping blocks of 2$\times$2 degree each, is consistent with zero in the radial range that we use in this work.\\


\subsubsection{Contamination \label{sec_contam}}
\hspace{4mm} A fraction of our source galaxies is physically associated with the clusters. They are not lensed and therefore dilute the lensing signal. We cannot remove them from the source sample because we lack redshifts. We could remove the bright early-type cluster members using their colours, but the faint cluster members cannot be efficiently removed because their red sequence is not well defined, and because many of them are blue \citep{Hoekstra07}. Fortunately, we can account for the dilution of the lensing signal by measuring the excess source galaxy density around the lenses, $f_{\rm cg}(r)$, and boost the lensing signal with 1+$f_{\rm cg}(r)$. As a further precaution, we exclude measurements on scales \mbox{$<$150 $h_{70}^{-1}$kpc} in our analysis. \\
\indent This correction implicitly assumes that the satellite galaxies are randomly oriented. If the satellites are preferentially radially aligned to the lens, however, the contamination correction may be too low. Most recent studies in this field report no significant radial alignment \citep[e.g.][]{Sheldon09I,Hao11,Schneider13,Chisari14,Sifon15}, although some earlier work claimed that such an alignment exists \citep[e.g.][]{Pereira05,Faltenbacher07}.  \\
\indent We do not account for the bias in the contamination correction which results from the blocking of the background sky by large (foreground) cluster galaxies \citep{simet15}. In \citet{VanUitert15}, we estimate it for LRGs in the RCS2 and find that the correction is biased by 5\% for low-redshift LRGs at a projected separation of \mbox{50 $h_{70}^{-1}$kpc}. For higher redshifts, and at larger separations, the bias quickly decreases, hence it is safe to neglect it here. \\
\indent Magnification by the clusters can also increase or decrease the source density, and hence bias the contamination correction. The ratio of the lensed and unlensed source counts (i.e. the bias) is given by $\mu^{\alpha - 1}$, with $\mu$ the lensing magnification and $\alpha$ the power law index of the flux number density distribution. Using the photometric redshift catalogue of \citet{Ilbert13}, we find that $\alpha=1.09$ at $22<r'<24$. To estimate the size of this bias, we assume that the mass distribution follows an NFW profile. For a cluster at $z=0.3$ with a mass of $M_{200}=5\times10^{14}h_{70}^{-1}M_\odot$ (the largest mass we probe), we find a bias of $\sim$5\% at 150 $h_{70}^{-1}$ kpc (the smallest lens-source separation we use). For lower masses and at larger separations, the bias becomes much smaller. This bias is smaller than our statistical errors and we can safely ignore it.


\subsection{Modelling of the signal \label{sec_hm}}
\hspace{4mm} In recent years it has become common practice to model the weak-lensing signal around galaxies and galaxy clusters using the halo model \citep{Seljak00,CoorayS02}. This model provides a statistical framework of the distribution of matter in the Universe. Basically, it assumes that this distribution can be modelled as a collection of separate dark matter haloes in which galaxies and galaxy clusters are embedded. The lensing signal comes from the haloes that host the galaxy cluster and from neighbouring dark matter haloes. \\
\indent The halo models we create are similar to those described in \citet{Johnston07}. Our model consists of four terms, namely the contribution of the BCG, $\Delta \Sigma_{\rm{BCG}}$, the contribution of the cluster halo, $\Delta \Sigma_{\rm{NFW}}$, the contribution from miscentred haloes, $\Delta \Sigma_{\rm{mis}}$, and a term that accounts for the contribution from neighbouring haloes, $\Delta \Sigma_{\rm{2h}}$. The models of \citet{Johnston07} also included a correction to account for the fact that the shear is not exactly linear. This term is only important on small scales for the most massive clusters. Since we only fit scales larger than \mbox{150 $h_{70}^{-1}$ kpc}, this term can be safely ignored.\\
\indent The lensing signal of the BCG is sufficiently accurately modelled as a point source. We model its mass using the BCG-halo mass scaling relation from \citet{Johnston07}: \mbox{$M_{\rm BCG}=1.334\cdot10^{12}/(1+[M_{200}/6.717\cdot10^{13}]^{-1.38})$}, in units of $h^{-1}M_\odot$. The contribution of the BCG to the total lensing signal is very small on scales $>150 h^{-1}_{70}$ and we merely add it for completeness.\\
\indent The central dark matter halo is described by a Navarro-Frenk-White profile \citep[NFW; ][]{Navarro96}. The NFW density profile is given by
\begin{equation}
  \rho(r) = \frac{\delta_c \rho_c}{(r/r_s)(1+r/r_s)^2},
\end{equation}
with $\delta_c$ the characteristic overdensity of the halo, $\rho_c$ the critical density for closure of the universe, and \mbox{$r_s=r_{200}/c_{\mathrm{200}}$} the scale radius, with $c_{\mathrm{200}}$ the concentration parameter. The NFW profile is fully specified for a given set of $(M_{200},c_{200})$, with $M_{200}$ the mass inside a sphere of radius $r_{200}$. We calculate the tangential shear profile using the analytical expressions provided by \citet{Bartelmann96} and \citet{WrightB00}. \\
\indent For a fraction of the clusters, the adopted BCG does not correspond to the actual centre of the dark matter distribution. To compute the lensing signal of a miscentred NFW profile, we first calculate the miscentred surface mass density:
\begin{equation}
\Sigma(r|r_{\rm mis})=\frac{1}{2\pi}\int_0^{2\pi} d\theta \Sigma_{\rm{NFW}} \big( \sqrt{r^2+r_{\rm mis}^2-2r r_{\rm mis} \cos(\theta)} \big)
\end{equation}
with $r_{\rm mis}$ the offset from the peak of the dark matter. Following \citet{Johnston07}, we assume that the miscentring distribution is reasonably well described by a 2D Gaussian,
\begin{equation}
P(r_{\rm mis})=\frac{r_{\rm mis}}{\sigma_s^2} \exp \left(-\frac{1}{2}(r_{\rm mis}/\sigma_s)^2 \right),
\end{equation}
with $\sigma_s$ the width of the distribution. For a given distribution, the mean surface mass density is then given by
\begin{equation}
\Sigma_{\rm{mis}}(r)=\int_0^{\infty} dr_{\rm mis} P(r_{\rm mis})\Sigma_{\rm{NFW}}(r|r_{\rm mis}).
\end{equation}
The model lensing signal is computed as usual using $\Delta\Sigma_{\rm{mis}}(r)=\overline{\Sigma}_{\rm{mis}}(<r)-\overline{\Sigma}_{\rm{mis}}(r)$.\\
\indent At large projected separations, neighbouring clusters significantly contribute to the lensing signal. This contribution is modelled as the two-halo term from the halo model presented in \citet{Mandelbaum06}, $\Delta \Sigma_{\rm{2h}}$. To avoid that neighbouring haloes overlap, we implement the halo-exclusion prescription as detailed in \citet{VanUitert11}. The amplitude of the two-halo term is set by the linear bias parameter, which is a free parameter in the fit. The main difference with this two-halo term compared to other commonly used descriptions \citep[e.g.][]{Cacciato09,VanUitert11,Leauthaud12} is that it is computed using the linear power spectrum rather than the non-linear one. We note that the regime where the one-halo and two-halo terms overlap (at a few Mpc) is difficult to model in general due to uncertainties in the prescription of halo exclusion and non-linear biasing. Marginalizing over the linear bias mitigates the impact of these uncertainties on the other fit parameters. We have implemented various variations of the two-halo term and we will discuss the effect this has on our results in the forthcoming sections. \\
\indent Close to massive clusters, the lensing signal is so large that the relation between the source galaxy ellipticities and the gravitational shear can no longer be approximated as being linear. To account for this, we convert the model shear to the reduced shear, $\Delta\Sigma^{\rm red}$, which is the quantity we measure observationally \citep{Seitz97,Hoekstra00,Applegate14}:
\begin{equation}
\Delta\Sigma^{\rm red}=\frac{\Delta\Sigma^{\rm model}}{1-\frac{\langle\beta_s^2\rangle}{\langle\beta_s\rangle^2}\frac{\Sigma^{\rm model}}{\Sigma_{\rm crit}}},
\label{eq_sigmared}
\end{equation}
where 
\begin{equation}
\beta_s=\frac{D_{\rm ls}}{D_{\rm s}}\frac{D_{\infty}}{D_{\rm l\infty}},
\end{equation}
with $D_{\infty}$ and ${D_{\rm l\infty}}$ the angular diameter distance from the observer to a source at infinity, and between the lens and a source at infinity. We determine $\langle\beta_s\rangle$ and $\langle\beta_s^2\rangle$ by integrating over the source redshift distribution, as determined from the photometric redshift catalogues of \citet{Ilbert13}. These reduced shear models are then fitted to the data. All lensing signals and model shears that we show in the following sections correspond to the reduced shear. Note that $\beta_s$ and $\Sigma_{\rm crit}$ in Equation \ref{eq_sigmared} are computed using the mean lens redshift for each cluster sample, rather than integrated over the lens redshift distribution. We have checked that the difference is of the order a few percent, and can be safely ignored. \\
\indent In short, the model that we fit to the lensing signal is given by
\begin{eqnarray}
\Delta \Sigma_{\rm{mod}}(r)=\Delta \Sigma_{\rm{BCG}}(r)+ p_c \Delta \Sigma_{\rm{NFW}}(r) +\nonumber \\
(1-p_c)\Delta \Sigma_{\rm{mis}}(r) + \Delta \Sigma_{\rm{2h}},
\end{eqnarray}
with $p_c$ the fraction of clusters that is correctly centred. Rather than $p_c$ we fit $q$, defined as $p_c\equiv 1/[1+\exp(-q)]$, which has an infinite range and can therefore be assigned a Gaussian prior. A high value of $q$ corresponds to a low miscentring fraction. \\


\subsubsection{Constraining the miscentring distribution \label{sec_crs}}
\indent Figure 6 in \citet{Johnston07} shows that the parameters that describe the miscentring distribution are not constrained by the lensing measurements. Their results are therefore sensitive to the adopted priors. If wrong priors are used, other parameters that are degenerate with the miscentring parameters may get biased, such as the concentration. To avoid such complications, we include the cluster-satellite correlation in the fit to obtain better constraints on the miscentring distribution, bypassing the need for using informative priors on the miscentring parameters. \\ 
\indent The satellites of a cluster trace the dark matter distribution, although the slope of their radial distribution may differ from the slope of the projected total mass distribution \citep[e.g.][]{Watson10,watson12,Budzynski12,Tal12}. Here, we use all red-sequence candidates at the cluster redshift brighter than $M^\star+1$ as satellites. We use this magnitude cut to ensure that our selection is homogeneous and complete over the entire redshift range of our clusters. We measure their radial distribution by correlating their positions to those of the BCGs. If a fraction of the clusters is not correctly centred, this also affects the observed distribution of satellites. The cluster-satellite correlation therefore provides additional constraints on the miscentring distribution of BCGs. Note that miscentring has a significantly smaller effect on $\Sigma$ than on $\Delta\Sigma$, which is illustrated in Fig. 4 in \citet{Johnston07}. However, the signal-to-noise ratio of the cluster-satellite correlation is five to ten times larger than the one from shear, so it still provides useful additional constraints. \\
\indent We measure the cluster-satellite correlation as follows:
\begin{equation}
\xi_{\rm cs}(r) = \bar{n}_{\rm rs} \left( \frac{N_{{\rm rs}}(r)}{N_{{\rm rand}}(r)}\times \frac{ N^{\rm tot}_{{\rm rand}}}{N^{\rm tot}_{{\rm rs}}} - 1 \right),
\end{equation}
with $N_{{\rm rs}}(r)$ and $N_{{\rm rand}}(r)$ the number of red-sequence galaxies and random points in a radial bin with a projected radial separation $r$ from the BCG. $N^{\rm tot}_{{\rm rand}}$ and $N^{\rm tot}_{{\rm rs}}$ are the total number of red-sequence galaxies and random points, respectively. The part between brackets measures the overdensity of red-sequence galaxies with respect to their average number density, $\bar{n}_{\rm rs}$. We therefore multiply it with $\bar{n}_{\rm rs}$ to convert it to the projected density of red-sequence galaxies in counts per Mpc$^{2}$. The signal of the total cluster sample is shown in Fig. \ref{plot_centdiff}. \\
\indent To account for zero-point and PSF variations, we scale the number of randoms to the number of red-sequence galaxies in each field separately. Secondly, we determine the ratio of the total number of red-sequence galaxies and the total number of random points in all fields as a function of position in the camera, and correct our measurement with this ratio. The purpose of this correction is two-fold. First, it ensures that the random points have exactly the same footprint as the red-sequence galaxies. Secondly, it accounts for PSF variations across the camera. In particular, the chips in the corners of the mosaic have fewer galaxies, as the average PSF is larger than in the central chips. \\
\indent To model the cluster-satellite correlation, we assume that the satellites trace the surface mass density of the dark matter, but, motivated by the results from \citet[][]{Watson10,watson12,Budzynski12,Tal12} and others, we allow the concentration of the satellites, $c_{\rm gal}$, to vary. Since we do not know a priori how the number of satellites is related to the surface mass density of the model, we fit this with a nuisance parameter, following \mbox{$\xi_{\rm cs}(r)=A_{\rm gal}\times\Sigma(r)$}, with $A_{\rm gal}$ in units $M^{-1}_\odot$, and marginalise over it. \\


\subsubsection{Intrinsic scatter mass-richness relation}
\begin{table}
  \caption{Priors on fit parameters}   
  \centering
  \begin{tabular}{c c c} 
  \hline
   & & \\
   Parameter & Prior-mean & Prior-sigma \\
   & & \\
  \hline\hline
   & & \\
 $\log_{10}(M_{200}$) $[h_{70}^{-1} M_\odot]$ & 14.15 & 2.00 \\
 $\ln(c_{200})$ & 1.39 & 3.00 \\
 $b$ & 2.62 & 4.00 \\
 $\ln(\sigma_s)$ [$h_{70}^{-1}$ Mpc] & -0.84 & 0.70 \\
 $q$ & 0.0 & 1.2 \\
 $\sigma_{\rm{ln}M|N_{200}}$ & see Table \ref{tab_clus} & 0.2\\
 $\ln(c_{\rm gal})$ & 1.39 & 3.00 \\
 $A_{\rm gal}[h_{70}M^{-1}_{\odot}]$ & 0.05 & 0.1 \\
  \hline \\
  \end{tabular}
  \label{tab_prior}
\end{table} 
\begin{figure}[t]
  \resizebox{\hsize}{!}{\includegraphics[angle=270]{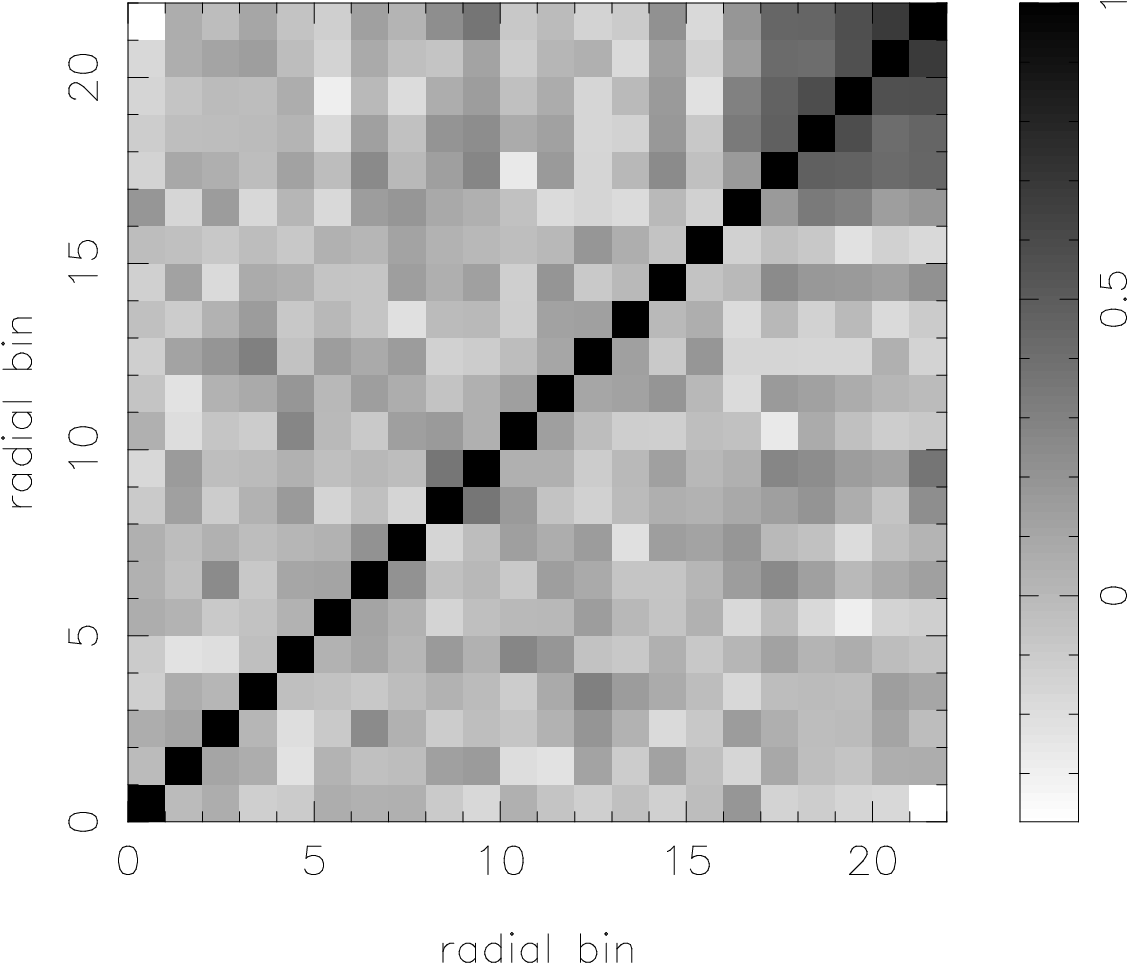}}
  \caption{Normalised covariance matrix between the radial bins of the lensing measurement and the cluster-satellite correlation for the clusters in the N4z1 sample (the fourth richness bin of the first redshift slice, see Table \ref{tab_clus}). The first 11 bins are the radial bins of the lensing measurements between \mbox{$0.15<r<5$ $h_{70}^{-1}$ Mpc}, the second set of 11 bins are the radial bins of the cluster-satellite correlation between \mbox{$0.15<r<5$ $h_{70}^{-1}$ Mpc}.}
  \label{plot_covar}
\end{figure}
\indent The mass-richness relation has intrinsic scatter; therefore, the best-fit lensing mass is not equal to the mean mass of the clusters in a bin. To account for this scatter, \citet{Johnston07} integrate their models over the probability distribution of halo masses, $P(M_{200})$, given by a log-normal distribution of $M_{200}$ for a given $N_{200}$, and fit the variance in ln($M_{200}$). The results of \citet{Becker07} and \citet{Evrard08} are used as a prior on the variance, which are based on a satellite kinematics study of maxBCG clusters and simulations, respectively. \\
\indent More recently, \citet{Rozo09a} studied the scatter in the mass-richness relation using X-ray measurements of the maxBCG clusters, and found $\sigma_{\rm{ln}M|N_{200}}=0.45_{-0.18}^{+0.20}$ (95\% confidence limits). This variance is smaller than the one from \citet{Becker07}; the difference was likely caused by the fact that \citet{Becker07} did not account for the miscentring of clusters. Therefore, we use the results from \citet{Rozo09a} to account for the scatter.\\
\indent Since the scatter cannot be constrained by our data, it is important to estimate the prior as accurately as possible. The scatter of the mass-richness relation from \citet{Rozo09a} is computed at a fixed richness. Our richness bins span a considerable range, which broadens the actual distribution of halo masses. To obtain a more representative value for the scatter, we take the following approach. We use a mass-richness relation to predict the mass of each cluster in a particular richness bin, and adopt that as the mean of a log-normal probability distribution with a constant $\sigma_{\rm{ln}M|N_{200}}=0.45$. We then combine the probability distributions of all clusters in that richness bin, and fit a log-normal distribution to it. We adopt the best-fit width as the scatter and list it in Table \ref{tab_clus}. We set the prior width to 0.2, which is the error on $\sigma_{\rm{ln}M|N_{200}}$ from \citet{Rozo09a}. \\
\indent We use the mass-richness relation from \citet{Johnston07}, \mbox{$M_{200}=8.8\times 10^{13} (N_{200}/20)^{1.28} h^{-1} M_{\odot}$}. In principle, we could follow an iterative approach where we update the mass-richness relation with our findings, but given that the relation we derive is not very different, and given that the derived scatter is fairly insensitive on the adopted relation, we regard this as unnecessary.
\begin{table*}
\renewcommand{\tabcolsep}{0.05cm}
  \caption{Cluster sample details}   
  \centering
  \begin{tabular}{c c c c c c c c c c c c c c} 
  \hline
   & & & & & & & & & & & & & \\
  bin & $N_{200}$ & n$_{\rm{clus}}$ & $\langle z \rangle$ & $\langle N_{200}\rangle$ &  $ N^{\rm{corr}}_{200}$ & $\sigma_{\rm ln M | N_{200}}$ & $M_{200}$ & $c_{200}$ & $b$ & $\sigma_s$ & $p_c$ & $c_{\rm gal}$ & $A_{\rm gal}$  \\
   & & & & & & & $[10^{13} h_{70}^{-1}M_{\odot}]$ & & & $[h_{70}^{-1}$ Mpc] & & &  [$\times 10^2h_{70}M^{-1}_\odot$] \\
   & (1) & (2) & (3) & (4) & (5) & (6) & (7) & (8) & (9) & (10) & (11) & (12) & (13) \\
   & & & & & & & & & & & & \\
  \hline\hline
   & & & & & & & & & & & & \\
\multicolumn{3}{l}{ $0.20<z<0.35$} \\
N1z1 & 2.00 - 7.63 & 534 & 0.30 & 5.2 & 4.0 & 0.57 & $5.0_{-0.6}^{+0.6}$ & $4.8_{-1.5}^{+2.5}$ & $2.6_{-0.4}^{+0.4}$ & $0.20_{-0.02}^{+0.03}$ & $0.39_{-0.12}^{+0.11}$ & $19.5_{-6.0}^{+13.8}$ & $2.8_{-0.3}^{+0.4}$ \\  
N2z1 & 7.63 - 10.47 & 345 & 0.28 & 9.0 & 7.5 & 0.47 & $8.9_{-0.9}^{+0.8}$ & $7.5_{-2.2}^{+3.9}$ & $3.1_{-0.4}^{+0.5}$ & $0.24_{-0.02}^{+0.02}$ & $0.21_{-0.06}^{+0.07}$ & $29.3_{-8.6}^{+15.7}$ & $3.2_{-0.3}^{+0.3}$ \\  
N3z1 & 10.47 - 13.65 & 268 & 0.27 & 11.9 & 10.4 & 0.46 & $7.7_{-0.9}^{+0.9}$ & $3.3_{-0.9}^{+1.4}$ & $2.5_{-0.4}^{+0.4}$ & $0.25_{-0.03}^{+0.03}$ & $0.48_{-0.09}^{+0.09}$ & $12.4_{-2.9}^{+4.8}$ & $5.0_{-0.4}^{+0.5}$ \\  
N4z1 & 13.65 - 18.14 & 216 & 0.27 & 15.8 & 14.3 & 0.46 & $11.8_{-1.3}^{+1.3}$ & $6.0_{-1.5}^{+2.1}$ & $3.1_{-0.4}^{+0.5}$ & $0.38_{-0.03}^{+0.03}$ & $0.37_{-0.05}^{+0.05}$ & $19.5_{-7.0}^{+15.2}$ & $4.9_{-0.4}^{+0.5}$ \\  
N5z1 & 18.14 - 26.31 & 197 & 0.25 & 21.8 & 21.2 & 0.47 & $13.8_{-1.5}^{+1.4}$ & $14.4_{-4.9}^{+9.1}$ & $3.8_{-0.5}^{+0.5}$ & $0.38_{-0.03}^{+0.03}$ & $0.30_{-0.05}^{+0.06}$ & $10.9_{-3.0}^{+4.8}$ & $5.9_{-0.4}^{+0.5}$ \\  
N6z1 & 26.31 - 220.00 & 114 & 0.24 & 38.0 & 36.6 & 0.57 & $29.3_{-3.0}^{+3.2}$ & $4.0_{-0.8}^{+1.1}$ & $2.8_{-0.9}^{+0.9}$ & $0.42_{-0.10}^{+0.12}$ & $0.59_{-0.10}^{+0.10}$ & $3.7_{-0.7}^{+0.9}$ & $4.6_{-0.3}^{+0.3}$ \\  
   & & & & & & & & & & & & & \\
\multicolumn{3}{l}{ $0.35<z<0.55$} \\
N1z2 & 2.00 - 7.63 & 1648 & 0.47 & 5.0 & 4.0 & 0.59 & $6.4_{-0.6}^{+0.6}$ & $12.7_{-4.1}^{+7.7}$ & $4.8_{-0.4}^{+0.4}$ & $0.20_{-0.01}^{+0.01}$ & $0.17_{-0.05}^{+0.06}$ & $9.3_{-1.2}^{+1.6}$ & $2.6_{-0.2}^{+0.2}$ \\  
N2z2 & 7.63 - 10.47 & 607 & 0.46 & 8.9 & 7.7 & 0.47 & $8.7_{-1.1}^{+1.0}$ & $12.7_{-4.8}^{+10.4}$ & $3.5_{-0.4}^{+0.5}$ & $0.28_{-0.02}^{+0.02}$ & $0.26_{-0.06}^{+0.05}$ & $28.6_{-7.4}^{+12.5}$ & $4.4_{-0.4}^{+0.5}$ \\  
N3z2 & 10.47 - 13.65 & 326 & 0.46 & 11.9 & 10.5 & 0.46 & $11.1_{-1.4}^{+1.5}$ & $4.1_{-1.4}^{+2.2}$ & $4.1_{-0.6}^{+0.6}$ & $0.27_{-0.03}^{+0.03}$ & $0.36_{-0.08}^{+0.07}$ & $10.8_{-2.4}^{+3.9}$ & $4.2_{-0.4}^{+0.5}$ \\  
N4z2 & 13.65 - 18.14 & 216 & 0.47 & 15.7 & 14.2 & 0.46 & $17.9_{-2.2}^{+2.3}$ & $9.3_{-3.0}^{+5.5}$ & $4.8_{-0.9}^{+0.9}$ & $0.31_{-0.03}^{+0.03}$ & $0.33_{-0.07}^{+0.07}$ & $10.9_{-2.3}^{+3.4}$ & $3.7_{-0.3}^{+0.4}$ \\  
N5z2 & 18.14 - 26.31 & 126 & 0.46 & 21.2 & 19.5 & 0.47 & $22.2_{-3.1}^{+3.0}$ & $14.1_{-5.4}^{+10.9}$ & $5.9_{-0.8}^{+0.9}$ & $0.40_{-0.03}^{+0.03}$ & $0.30_{-0.05}^{+0.05}$ & $19.6_{-6.0}^{+11.0}$ & $4.5_{-0.5}^{+0.6}$ \\  
N6z2 & 26.31 - 220.00 & 59 & 0.46 & 36.9 & 34.0 & 0.56 & $41.0_{-5.9}^{+6.1}$ & $6.5_{-2.4}^{+4.2}$ & $5.1_{-1.0}^{+1.0}$ & $0.41_{-0.05}^{+0.05}$ & $0.32_{-0.07}^{+0.08}$ & $11.9_{-3.1}^{+4.9}$ & $3.8_{-0.4}^{+0.5}$ \\  
   & & & & & & & & & & & & & \\
\multicolumn{3}{l}{ $0.55<z<0.80$} \\
N1z3 & 2.00 - 7.63 & 1170 & 0.64 & 5.1 & 3.3 & 0.58 & $4.9_{-0.8}^{+0.8}$ & $10.4_{-5.1}^{+14.3}$ & $6.3_{-0.8}^{+0.9}$ & $0.23_{-0.02}^{+0.03}$ & $0.35_{-0.09}^{+0.09}$ & $11.6_{-2.7}^{+4.7}$ & $3.0_{-0.4}^{+0.5}$ \\  
N2z3 & 7.63 - 10.47 & 670 & 0.66 & 9.0 & 6.6 & 0.46 & $6.7_{-1.3}^{+1.3}$ & $45.9_{-30.4}^{+95.1}$ & $4.3_{-0.8}^{+0.8}$ & $0.28_{-0.01}^{+0.01}$ & $0.23_{-0.05}^{+0.05}$ & $152.5_{-60.5}^{+104.1}$ & $4.8_{-0.7}^{+1.0}$ \\  
N3z3 & 10.47 - 13.65 & 512 & 0.67 & 11.9 & 9.4 & 0.46 & $11.0_{-1.7}^{+1.8}$ & $35.7_{-20.7}^{+68.0}$ & $5.5_{-0.8}^{+0.9}$ & $0.30_{-0.02}^{+0.02}$ & $0.22_{-0.05}^{+0.05}$ & $36.1_{-11.4}^{+23.2}$ & $3.7_{-0.5}^{+0.6}$ \\  
N4z3 & 13.65 - 18.14 & 340 & 0.67 & 15.5 & 13.0 & 0.46 & $13.7_{-2.8}^{+3.0}$ & $3.4_{-1.6}^{+3.3}$ & $5.5_{-1.0}^{+1.2}$ & $0.34_{-0.02}^{+0.02}$ & $0.29_{-0.05}^{+0.05}$ & $51.7_{-19.3}^{+42.3}$ & $3.9_{-0.7}^{+0.9}$ \\  
N5z3 & 18.14 - 26.31 & 190 & 0.67 & 21.3 & 20.1 & 0.47 & $18.1_{-3.7}^{+3.7}$ & $10.3_{-5.3}^{+12.6}$ & $6.4_{-1.2}^{+1.2}$ & $0.39_{-0.02}^{+0.01}$ & $0.31_{-0.05}^{+0.05}$ & $135.8_{-67.0}^{+112.8}$ & $4.4_{-0.7}^{+1.0}$ \\  
N6z3 & 26.31 - 220.00 & 73 & 0.68 & 36.0 & 33.1 & 0.55 & $29.2_{-7.2}^{+7.9}$ & $4.7_{-2.5}^{+6.5}$ & $6.7_{-1.4}^{+1.6}$ & $0.33_{-0.06}^{+0.05}$ & $0.44_{-0.10}^{+0.12}$ & $7.6_{-1.9}^{+3.2}$ & $3.5_{-0.6}^{+0.9}$ \\  
  \hline \\
  \end{tabular}
  \label{tab_clus}
\tablefoot{(1) richness range of the bin; (2) number of clusters; (3) median redshift; (4) mean richness; (5) mean richness corrected for Eddington bias; (6) prior-mean of the scatter in the mass-richness relation; (7) halo mass; (8) concentration; (9) halo bias; (10) miscentring width; (11)  centring fraction; (12) concentration of satellite distribution; (13) amplitude of cluster-satellite correlation.}
\end{table*} 
\begin{figure*}
  \resizebox{\hsize}{!}{\includegraphics{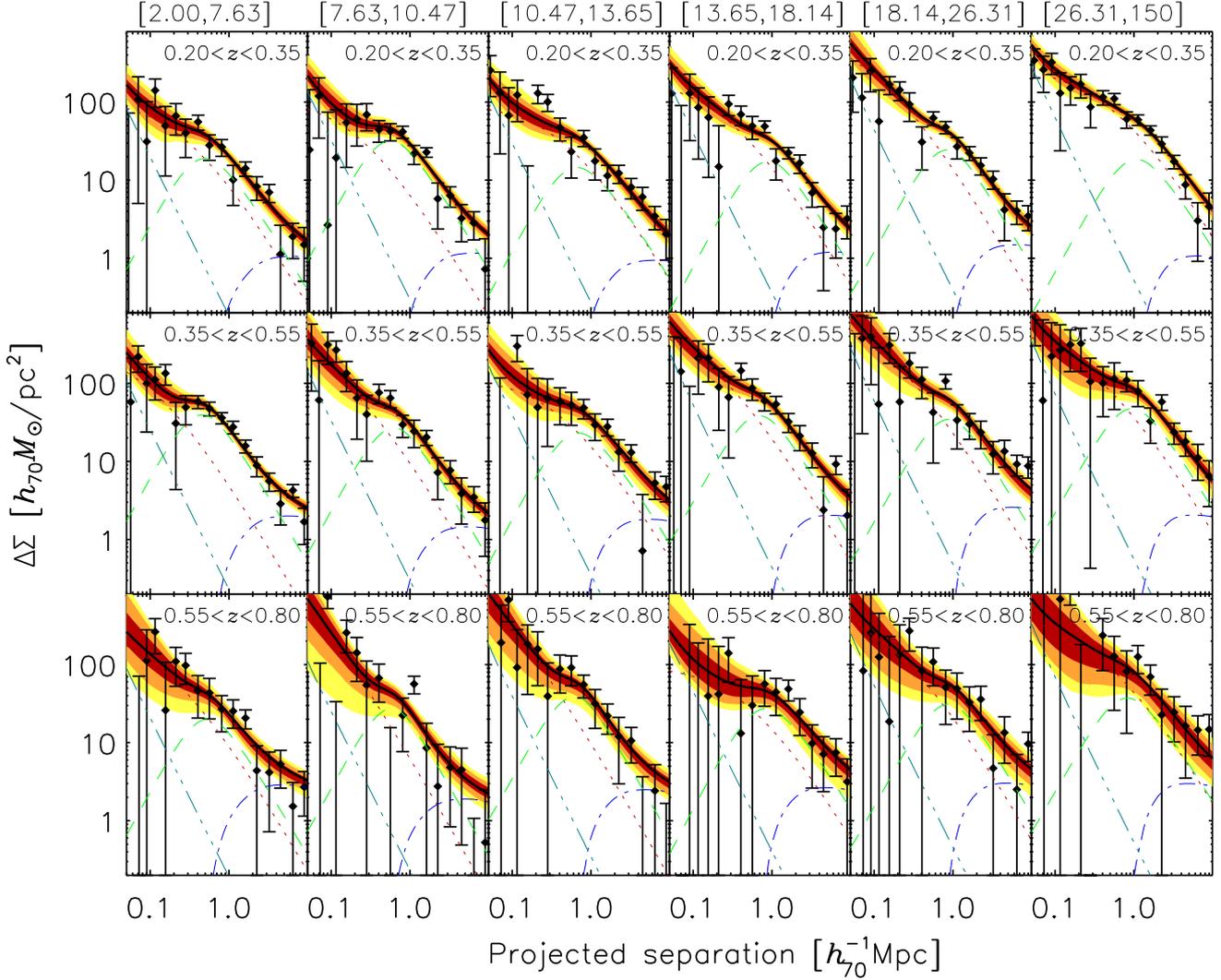}}
  \caption{Lensing signal $\Delta \Sigma$ as a function of projected separation from the BCG for the different cluster samples, split in richness (as indicated on top of each column) and redshift (indicated in each panel). The solid black lines indicate the best-fit cluster halo model, simultaneously fitted to the lensing signal and the corresponding cluster-satellite correlation signal in the range $0.15<r<5$ $h_{70}^{-1}$ Mpc. The dotted-dotted-dashed grey line shows the contribution from the BCG, the dotted red line the contribution from correctly centred clusters, the dashed green line the contribution from miscentred clusters, and the dotted-dashed blue line the contribution from neighbouring haloes. The dark red, orange and yellow shaded areas correspond to the 1-, 2- and 3-$\sigma$ model uncertainty.}
  \label{plot_gg_n200}
\end{figure*}


\subsubsection{Model fitting}
\begin{figure*}
  \resizebox{\hsize}{!}{\includegraphics{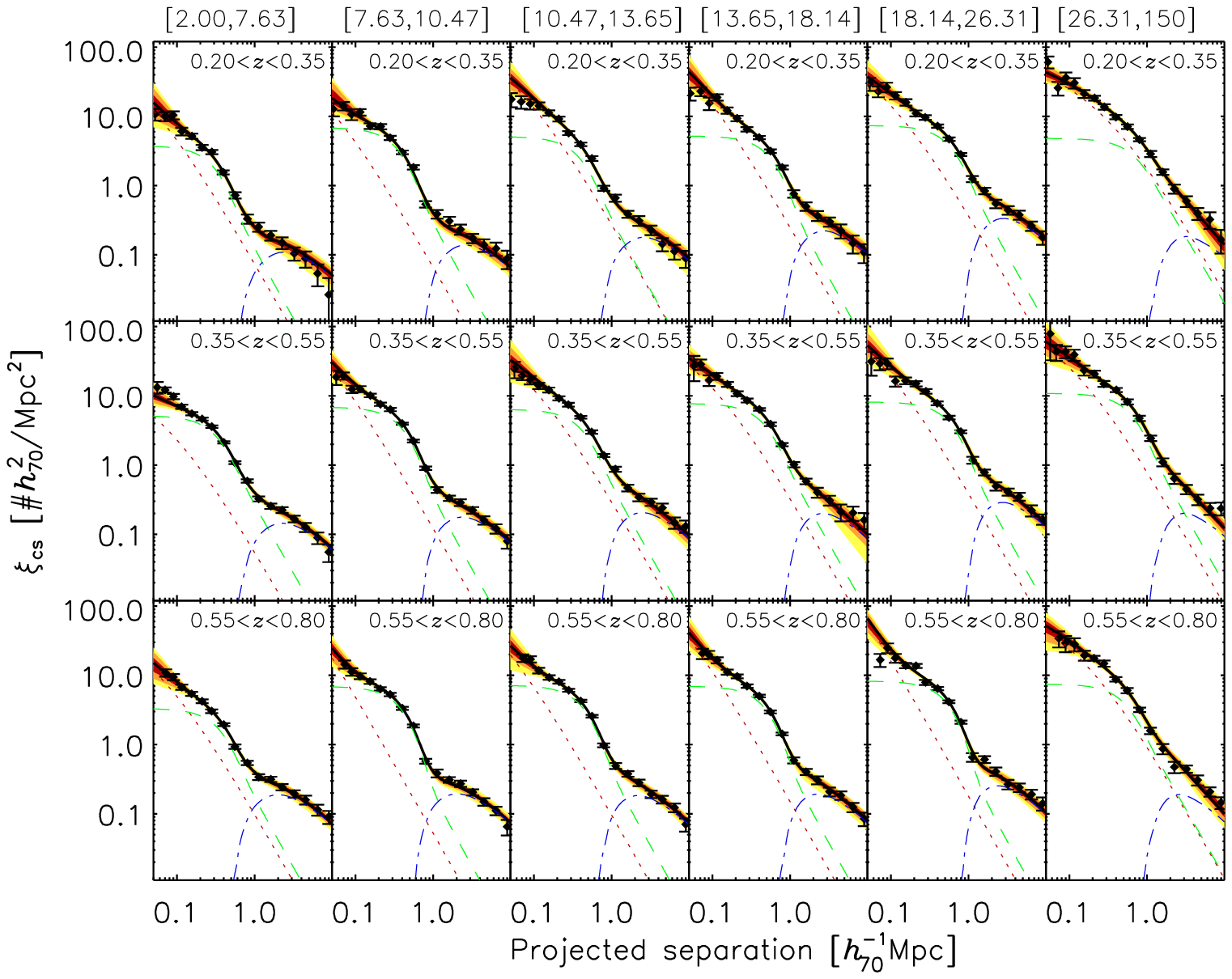}}
  \caption{Cluster-satellite correlation signal, measured using red-sequence candidates brighter than $M^\star+1$, as a function of projected separation from the BCG, for the different cluster samples, split in richness (as indicated on top of each column) and redshift (indicated in each panel). The solid black lines indicate the best-fit cluster halo models, obtained from simultaneous fits to the lensing signal and the cluster-satellite correlation signal in the range $0.15<r<5$ $h_{70}^{-1}$ Mpc. The dotted red line shows the contribution from correctly centred clusters, the dashed green line the contribution from miscentred clusters, and the dotted-dashed blue line the contribution from neighbouring haloes. Note that the measurements are correlated at large scale (see Fig. \ref{plot_centdiff}), which is accounted for in the fit. The dark red, orange and yellow shaded areas correspond to the 1-, 2- and 3-$\sigma$ model uncertainty.}
  \label{plot_galod_all}
\end{figure*}
\indent To efficiently sample parameter space and fit the models in a reasonable amount of time, we use {\sc Emcee} \citep{Foreman13}, the publicly available ensemble Markov Chain Monte Carlo (MCMC) sampler. The free parameters in this model are $M_{200}$, $c_{\rm{200}}$, $b$, $\sigma_s$, $q$, $\sigma_{\rm{ln}M|N_{200}}$, $c_{\rm gal}$ and $A_{\rm gal}$. For convenience, we summarise the priors in Table \ref{tab_prior}. Note that for all but one parameter, $\sigma_{\rm{ln}M|N_{200}}$, the priors are uninformative, hence the results do not depend on them. We assess how the best-fit masses depend on the prior of $\sigma_{\rm{ln}M|N_{200}}$ in the results section. \\
\indent We run {\sc Emcee} with 200 walkers, starting near the best-fit model of each sample. The number of steps of each walker is set to 3 000. We conservatively discard the first 500 steps as the burn-in phase. The resulting \mbox{500 000} model evaluations are used to estimate the parameter uncertainties. The fit parameters and their errors which we report in the following sections correspond to the median and the 68\% confidence intervals of the marginalised posterior distributions. \\
\indent To fit the models to the data, we need to estimate the covariance between the data points. We do this by measuring the signal in each of our 145 2$\times$2 degree patches, from which we create a large number of bootstrap realisations. The covariance matrices are estimated from the different realisations. We show a representative normalised covariance matrix for one of our cluster samples in Fig. \ref{plot_covar}. This figure shows that the lensing measurements are not correlated. We therefore set the off-diagonal elements of the lensing signal, and the correlation of the lensing signal with the cluster-satellite correlation, to zero. The cluster-satellite correlation itself, however, is correlated at large scales, hence we keep those off-diagonal terms. The inverse of the resulting covariance matrix is used in the fit. To correct for the bias which is introduced when a noisy covariance matrix is inverted \citep{Hartlap07}, we multiply the inverse covariance matrix with a correction factor $(N_{\rm sample}-N_{\rm data}-2)/(N_{\rm sample}-1)$, where $N_{\rm sample}$ is the number of independent samples used in the bootstrap, in our case 145, and $N_{\rm data}$ is the number of data points, 22.\\


\subsection{Fit results}
\indent We divide the clusters in bins of richness and redshift, as detailed in Table \ref{tab_clus}. Although inherently somewhat arbitrary, these ranges were chosen such that they enable us to reliably measure and study potential trends with richness and redshift.
The stacked lensing signals are shown in Fig. \ref{plot_gg_n200} and the cluster-satellite correlations in Fig. \ref{plot_galod_all}, together with the best-fit halo models. The errors on the measurements correspond to the square root of the diagonal of the covariance matrix and indicate the 68\% confidence intervals. The trends in the data, such as the kink at $\sim$1 $h_{70}^{-1}$ Mpc due to the miscentring of clusters, are well reproduced by the model. We find an average reduced chi-squared value of $\chi^2_{\rm red}=1.1$ (with 14 d.o.f. per bin), suggesting that the data is well modelled. When we average the residuals of the fit for all bins, we find that the models underpredict the cluster-satellite correlation at scales 1-2 $h_{70}^{-1}$Mpc, exactly the regime that is difficult to model because of the overlap between the one-halo and two-halo term. \\
\begin{figure*}
  \center
  \resizebox{0.81\hsize}{!}{\includegraphics{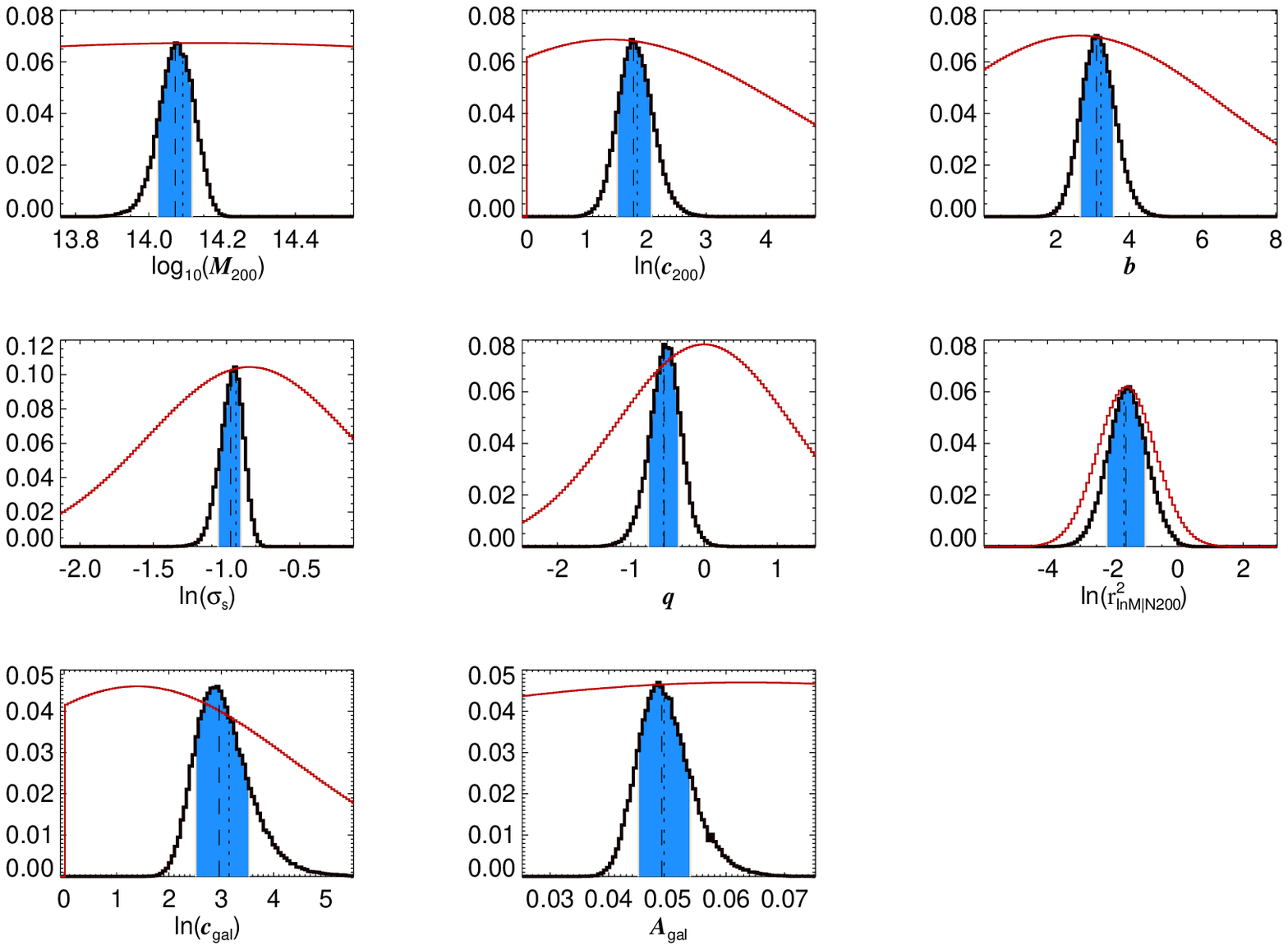}}
  \caption{Posteriors of the fitted cluster halo model parameters for the N4z1 bin, marginalised over all other parameters. Black solid lines indicate the posterior, red lines the prior. The dashed vertical line indicates the median of the marginalised posterior, the blue shaded area is the 68\% confidence interval around the median, and the dotted vertical line indicates the location of the best-fit value. Dimensions as in Table \ref{tab_clus}.}
  \label{plot_posterior}
\end{figure*}
\begin{figure*}
  \center
  \resizebox{0.81\hsize}{!}{\includegraphics{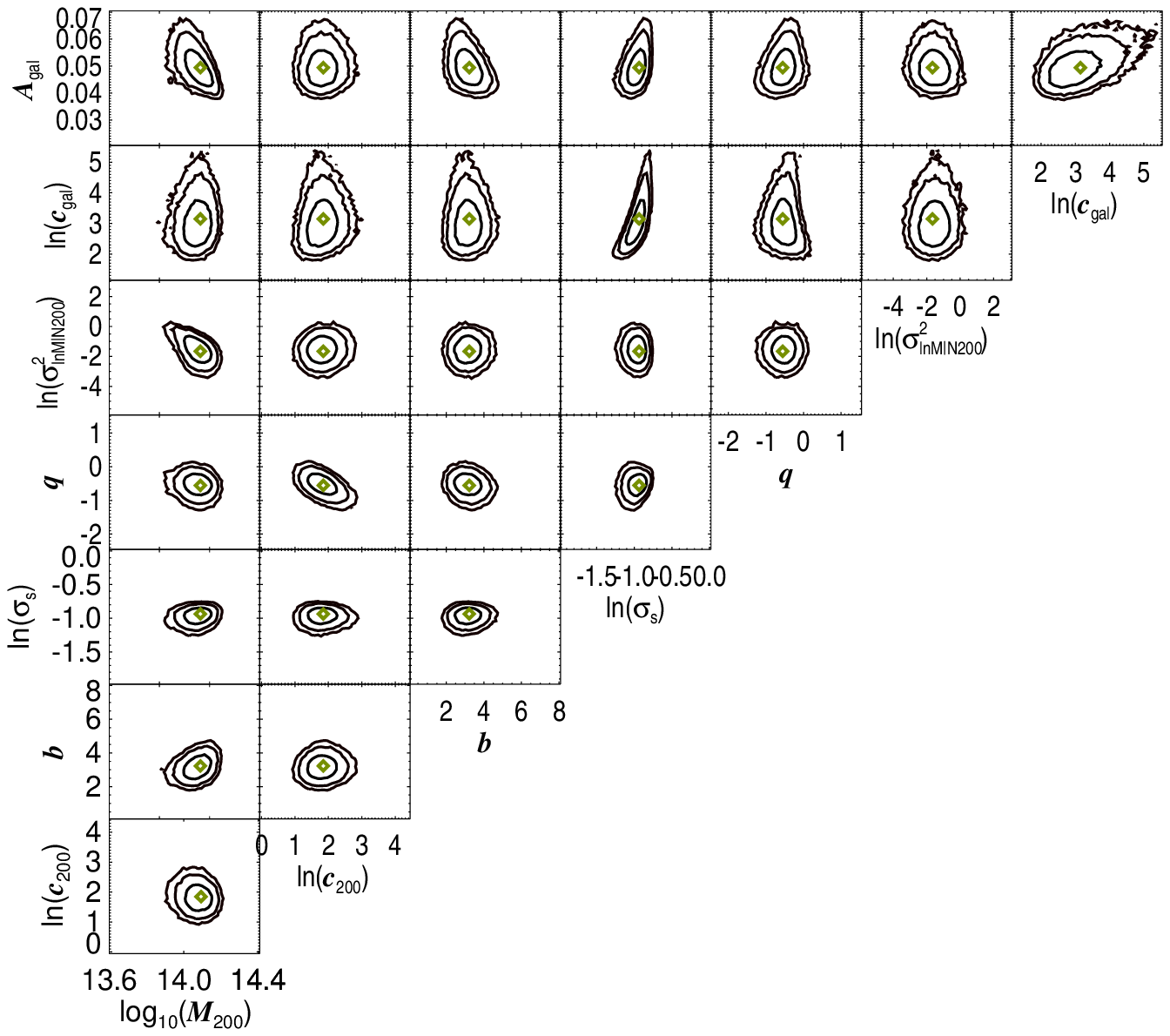}} 
  \caption{Posteriors of the fitted cluster halo model parameters for the N4z1 bin, for all pairs of parameters. Shown are the 1$\sigma$, 2$\sigma$ and 3$\sigma$ confidence regions. The best-fit values are indicated by the green open diamonds. These plots illustrate the degeneracies that exist between the fit parameters. Dimensions as in Table \ref{tab_clus}.}
  \label{plot_posterior2}
\end{figure*}
\indent To illustrate how well the model parameters are constrained, we show the marginalised posteriors of the fit parameters for the N4z1 bin in Fig. \ref{plot_posterior}, together with the priors that were used in the fit. Only $\sigma_{\rm{ln}M|N_{200}}$ cannot be constrained by the data. For the other parameters, the choice of the prior is not important as they are well constrained. A wrong choice for the prior of $\sigma_{\rm{ln}M|N_{200}}$, however, could bias our results if degeneracies exist. To investigate this, we show the two-dimensional marginalised posteriors of all pairs of parameters in Fig. \ref{plot_posterior2}. $\sigma_{\rm{ln}M|N_{200}}$ is only somewhat degenerate with $M_{200}$, but not with the other parameters. To assess the sensitivity of our results to the chosen priors, and to enable a more detailed comparison to the results of \citet{Johnston07}, we also fit our cluster halo models but {\it only} to the shear, adopting the priors that were used in \citet{Johnston07}. We discuss how that affects our results in the following section. \\


\section{Mass-richness relation \label{sec_mr}}
\indent In order to enable a comparison of our mass-richness relation to simulations, we have to account for Eddington bias: the observational scatter which causes clusters to preferentially move from richness ranges where the abundance of clusters is high to where it is low. This is a separate effect from intrinsic scatter, which defines the width of the halo mass distribution at a given richness if both quantities could be measured with infinite precision, which we account for in our halo model. The observational scatter is mainly caused by the field-to-field variance of the background number density of red-sequence galaxies, for which we use a global estimate. We correct $N_{200}$ for Eddington bias using Bayes theorem. The probability distribution of the underlying $N_{200}$ given an observed value $N_{200}^{\rm{obs}}$ (the posterior) is proportional to the product of the chance of having a value of $N_{200}^{\rm{obs}}$ given a distribution of $N_{200}$ (likelihood) and the probability distribution of $N_{200}$ (prior):
\begin{equation}
p(N_{200}|N_{200}^{\rm{obs}})\propto p(N_{200}^{\rm{obs}}|N_{200})p(N_{200}).
\end{equation}
The likelihood is determined by the measurement uncertainty of $N_{200}$, which our cluster finder provides. The errors are $\sim$20\% larger than Poisson, but we adopt a Poisson distribution as a reasonable first order approximation for the likelihood distribution. For the prior, we could in principle use the observed richness distribution. However, as shown in \mbox{Fig. \ref{plot_z_n200}}, the cluster sample is incomplete at the low richness end by an uncertain amount, and using it as a prior would lead to an erroneous correction. Since we expect that the cluster sample is complete for approximately $N_{200}>15$, we fit a power law to the richness distribution at $20<N_{200}<40$. For the prior, we replace the observed richness distribution with this power law at $N_{200}<20$, whilst at larger richnesses we use the observed richness distribution. We sum the posteriors of all clusters in a bin, normalise it and integrate up to the mean, $N_{200}^{\rm{corr}}$. These values are tabulated in Table \ref{tab_clus}, as well as the uncorrected values. Only the corrected richnesses are used in the following, unless explicitly mentioned otherwise.  \\
\begin{figure}
  \resizebox{\hsize}{!}{\includegraphics{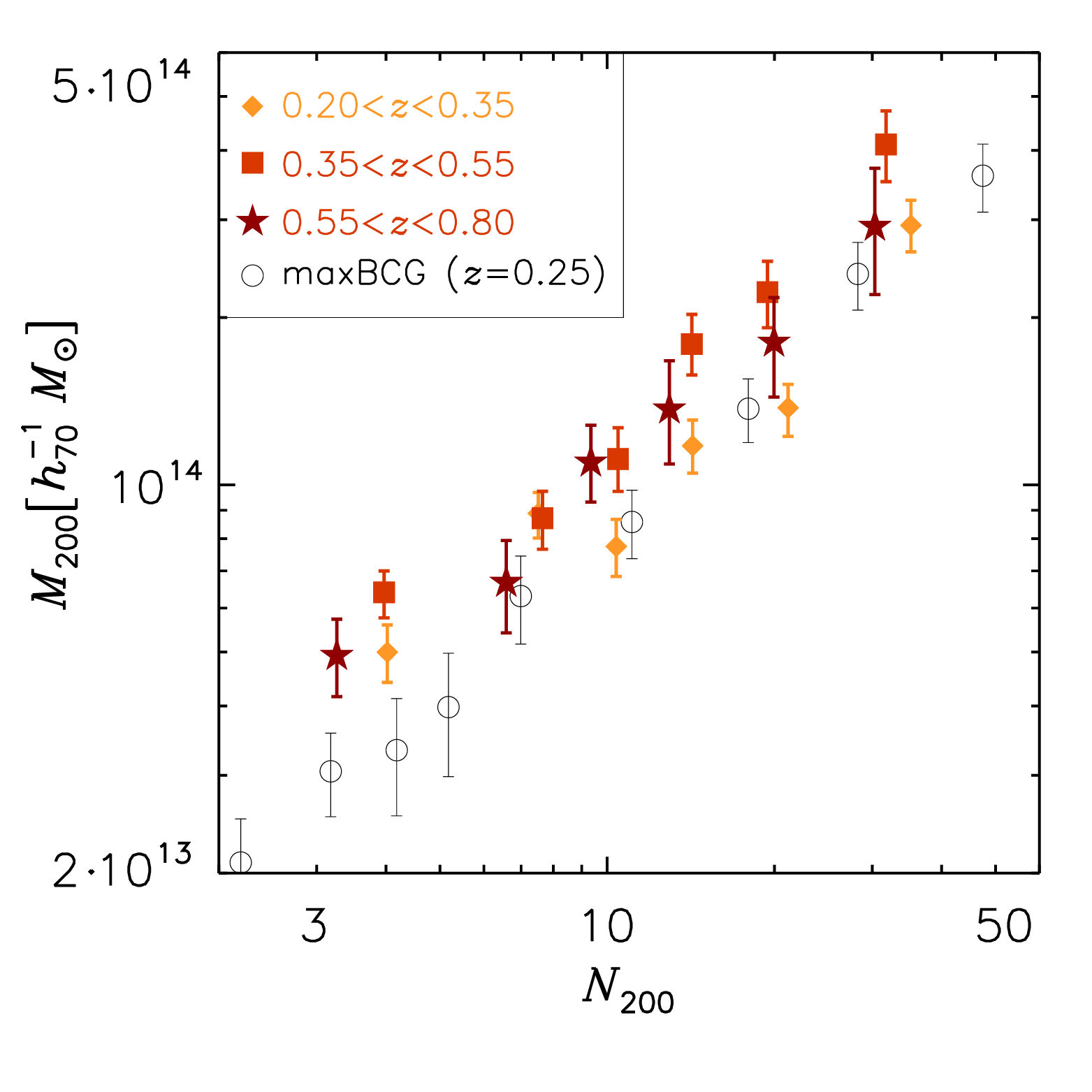}}
  \caption{Cluster mass versus richness for the cluster samples as indicated in the plot. The red circles correspond the results from the lensing analysis of the maxBCG clusters of \citet{Johnston07}, which have been boosted by a factor 1.18 to account for the impact of photometric redshift scatter (see text). All the measurements that are shown here have been corrected for Eddington bias.}
  \label{plot_n200m200}
\end{figure}
\indent We show the mass-richness relation in Fig. \ref{plot_n200m200}. Note that the mass we show is the mean of the log-normal distribution of halo masses we integrate over in our halo model. We fit a power-law relation of the form \mbox{$M_{200}=A(N^{\rm{corr}}_{200}/20)^{\alpha}$} in each redshift slice and report the best-fit slopes and normalisations in Table \ref{tab_n200m200}. The errors on the amplitude are determined by marginalising over the slope, and vice versa. The likelihood contours of the fit are shown in Fig. \ref{plot_like}. Without the correction for Eddington bias, we would have obtained $A=(14.4\pm0.7)\times10^{13} h_{70}^{-1} M_{\odot}$ and $\alpha=0.83\pm0.08$ for the 0.20$<$$z$$<0.35$ bin, which deviates by approximately 1$\sigma$.\\
\begin{figure}
  \resizebox{\hsize}{!}{\includegraphics[angle=270]{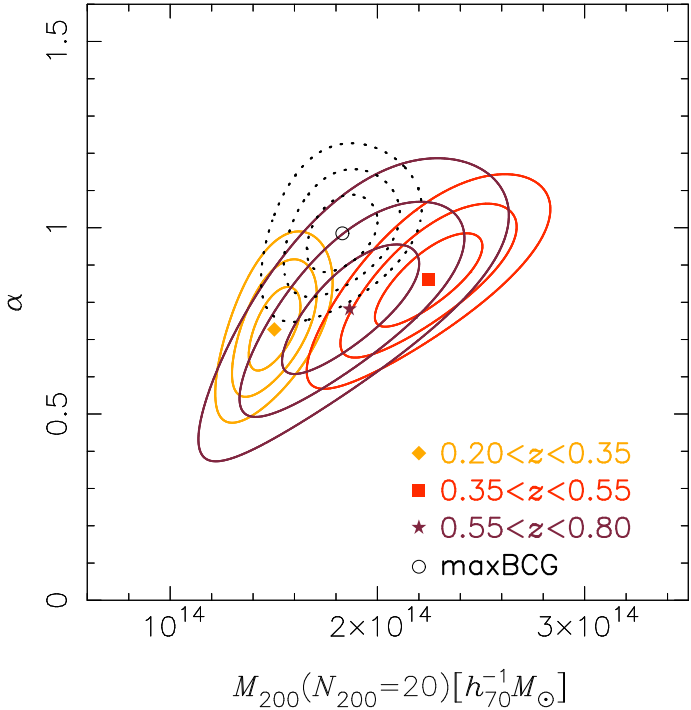}}
  \caption{67.8\%, 95.4\% and 99.7\% confidence limits of the fits to the mass-richness relation. The black dotted lines show the results from \citet{Johnston07} in the overlapping richness range.}
  \label{plot_like}
\end{figure}
\begin{table}
  \caption{Power-law parameters of the fit between $N_{200}$ and $M_{200}$.}   
  \centering
  \begin{tabular}{c c c } 
  \hline
   & & \\
 Bin & $A$ & $\alpha$  \\
   & [$10^{13} h_{70}^{-1} M_{\odot}$] & \\
   & & \\
  \hline\hline
   & & \\
0.20$<$$z$$<$0.35 & $15.0\pm0.8$ & $0.73\pm0.07$ \\
0.35$<$$z$$<$0.55 & $22.5\pm1.7$ & $0.86\pm0.08$ \\
0.55$<$$z$$<$0.80 & $18.7\pm2.2$ & $0.78\pm0.11$ \\
   & & \\
maxBCG & $18.3\pm 1.1$& $0.98\pm0.07$ \\
(3$<$$N_{200}$$<$50) & & \\
  \hline \\
  \end{tabular}
  \label{tab_n200m200}
\end{table} 
\indent A number of complications limit a simple interpretation of the weak lensing mass estimates of clusters, such as intrinsic profile variations of dark matter haloes \citep[e.g.][]{Clowe04,Corless07,Gruen15} and the presence of correlated and uncorrelated structure along the line-of-sight \citep[e.g.][]{Metzler01,Hoekstra01,Hoekstra11}. These complications mainly increase the scatter of the mass estimates, but may even lead to small ($\sim$5-10\%) biases if model fitting techniques are used \citep{Becker11,Rasia12}. The lensing signal can be modelled in various ways, and particular choices can reduce this bias \citep{Mandelbaum10}. More detailed numerical simulations are required to quantify this bias more precisely, e.g. as a function of mass and redshift, to interpret the results correctly. This is important for the exploitation of clusters as a reliable tool for cosmology.


\subsection{Comparison to previous single-redshift results}
\indent We compare our results to the weak lensing analysis of the maxBCG cluster sample \citep{Koester07}, a catalogue of 13\hspace{0.5mm}823 clusters that has been detected in the SDSS. The cluster detection algorithm employed in \citet{Koester07} identifies the cluster red-sequence galaxies, and selects the brightest, the BCG, as centre of the cluster. The resulting cluster sample covers the richness range $10<N_{200}<190$ and a redshift range of $0.1<z<0.3$. In \citet{Sheldon09I}, the cluster sample is extended to $N_{200}=3$, which leads to a sample of $\sim$130\hspace{0.5mm}000 galaxy groups and clusters. The lensing analysis of the sample is presented in \citet{Sheldon09I}; the mass-richness relation is derived in \citet{Johnston07}. Following \citet{Mandelbaum08}, we multiply the masses from \citet{Johnston07} by a factor 1.18 to account for the impact of photometric redshift scatter in the lensing analysis (as was done in \citet{Rozo10}). Note that \citet{Johnston07} use the same definition of mass as we do. \\
\indent The richnesses of \citet{Johnston07} have not been corrected for Eddington bias. We perform the correction, using a probability distribution for the maxBCG sample of $p(N_{200})\propto (N_{200})^{-3}$ over the entire richness range, following \citet{Andreon10}, which is in good agreement with a slope of $-3.06$ that we find for our clusters with $0.2<z<0.35$ and $N_{200}$$>$15. We adopt the mean richness as $N_{200}^{\rm obs}$ to compute the posterior, instead of stacking the posteriors of the individual clusters. This makes a negligible difference. We show the corrected results in Fig. \ref{plot_n200m200}. We fit the same power law in the overlapping richness range, $3<N^{\rm{corr}}_{200}<50$, and list the best-fit parameters in Table \ref{tab_n200m200}. The amplitude and slope of our low-redshift sample are about 3$\sigma$ lower than those of the maxBCG. Figure \ref{plot_n200m200} shows that this discrepancy is partly driven by the bins at $N_{200}$$<$10. If we fit the relation at $N_{200}$$>$10, the slopes are consistent but the amplitudes still differ by $\sim$2.5$\sigma$. \\
\indent There are several differences between the analyses. For instance, Fig. \ref{plot_richcomp} suggests that our richnesses are somewhat larger than those of the maxcBCG for rich systems; how they differ at low $N_{200}$ is unclear as the public maxBCG catalogue only includes $N_{200}$$>$10 clusters. If our richnesses are systematically larger than maxBCG for richer systems, this would tend to lower our normalisation and slope, which may partly explain the discrepancy. The number of matching RCS2 and maxBCG clusters is too low to assess this quantitatively. Also the purity of the two cluster samples may differ, as different cluster detection algorithms have been used on different data. \citet{Koester07} show how the purity of the sample depends on particular settings of the maxBCG algorithm using mock catalogues. For maxBCG, the purity is typically of the order 90\% or higher at richnesses $N_{200}>10$; how the purity varies at lower richnesses is not reported. Since the purity of the RCS2 cluster sample is expected to be high as well, it seems unlikely that differences in the purity  could lead to differences larger than a few percent in the masses. \\
\indent Also on the modelling side, there are noticeable differences. For example, we use different priors for the miscentring distribution and for the scatter between richness and halo mass. In \citet{Johnston07}, the miscentring priors are based on numerical simulations, from which a functional form is derived for $q$, with a corresponding $p_c$ (the fraction of clusters that is correctly centred) ranging from 60\% to 80\% for richnesses in the range of $N_{200}=10$ to $N_{200}=100$. The width of the miscentring distribution is fixed at \mbox{0.42$h^{-1}$Mpc}. Our best-fit parameter suggest a $p_c$ in the range of 20-50\%, and a narrower width \mbox{$\sigma_s \approx$ 0.2-0.4 $h_{70}^{-1}$ Mpc}. It is conceivable that the miscentring distribution differs for the two catalogues, as different algorithms have been applied to different data to identify the BCG. However, if the actual miscentring distributions are similar, the different priors could cause discrepancies between the best-fit masses and concentrations due to the parameter degeneracies. The same holds for the adopted prior for the scatter in mass-richness. Motivated by this, we run our cluster halo models on the lensing data only, using the priors from \citet{Johnston07}. Note that we do not account for the differences between the two-halo terms. Nevertheless, this test will give us a reasonable impression how sensitive our results are to the adopted priors. \\
\indent The resulting best-fit masses of the individual bins are consistently lower but within the 1$\sigma$ error bars of the nominal results. When we refit the power-law relation, we find an amplitude and slope of $A=(12.2\pm1.2)\times10^{13} h_{70}^{-1} M_{\odot}$ and $\alpha=0.69\pm0.15$ for the 0.20$<$$z$$<$0.35 redshift slice, which is even more discrepant. Comparing this to our nominal results shows that the tighter constraints on the miscentring distribution, obtained from including the cluster-satellite correlation in the fit, also leads to smaller errors on the mass. As the amplitude is $\sim$2$\sigma$ lower than our nominal result and the slope is consistent, our results do not critically depend on the adopted priors. Furthermore, we have also checked that by changing the implementation of the two-halo term, the masses do not change significantly. \\
\indent The mass-richness relation has also been derived from semi-analytic galaxy formation models based on N-body simulations. For example, \citet{Hilbert10} used the Millennium Simulation \citep{Springel05} and found that the derived mass-richness relation agreed well in shape and amplitude with the relation from maxBCG. \citet{Hilbert10} report an amplitude and slope of  \mbox{$A=(17.7\pm0.1)\times10^{13} h_{70}^{-1} M_{\odot}$} and $\alpha=1.09\pm0.01$. The amplitude and slope are a bit higher than our low-redshift results. The difference is again mainly driven by the $N_{200}$$<$10 results, but additionally, some difference may be caused by the different definitions of richness. \citet{Angulo12} measure the relation between optical richness and mass at $z$=0.25 in the Millennium-XXL simulation, which extends the Millennium and Millennium-II Simulations \citep{Springel05,Boylan-Kolchin09}. A power-law slope of 1.07 is reported, steeper than our results, but again the difference may originate from the low-richness end and from differences between the richness estimators. \\
\indent In \citet{Ford15}, a sample of 18 000 optically selected clusters at 0.2$<$$z$$<$0.9 are studied in the Canada-France-Hawai Telescope Lensing Survey \citep[CFHTLenS;][]{Heymans12}. Clusters are identified using the 3D-Matched-Filter finder of \citet{Milkeraitis10}. Richnesses are defined as all cluster members within $r_{200}$ that are brighter than an absolute $i$-band magnitude of $-19.35$, as detailed in \citet{Ford14}. It is not clear how that relates to our richness estimate. Using the same parametrisation of the mass-richness relation as we do, they find an amplitude and slope of $2.7^{+0.5}_{-0.4}\times10^{13}M_\odot$ and $1.4\pm0.1$, respectively. No correction for the Eddington bias is performed, leading to a larger slope. However, the normalisation is significantly lower than what we find. The difference may be partly attributed to the definition of richness (see the discussion in Sect 5.3 of \citet{Ford15}). 


\subsection{Redshift evolution}
\indent At any given richness, clusters at 0.35$<$$z$$<$0.55 have larger masses than those at 0.2$<$$z$$<$0.35. Whether this trend continues towards higher redshifts is not clear given the large errors on the mass for the 0.55$<$$z$$<$0.8 clusters, although our results suggest that it does not. \\
\indent Both the halo masses and the richnesses of clusters may evolve. The bulk of the change in halo mass is not expected to be physical accretion, but rather an effect called pseudo evolution \citep{Diemer13}. Halo masses are defined with respect to a background density (usually the mean or the critical density), which evolves with redshift. Even if a halo does not accrete anything, its mass increases with time as the background density drops. If the richness of a cluster would not change, we would expect to see an increase in halo mass towards lower redshifts. The fact that we find the opposite trend and that the halo mass decreases at a given richness, points towards an evolution of the richness.\\
\indent Clusters build up their red sequence over cosmic time. The cluster galaxies are stripped of their gas through tidal interactions and ram pressure stripping, which quenches their star formation \citep[e.g.][]{Boselli06}. Consequently, the late-type spiral galaxies that are accreted turn into early-type S0 galaxies, and subsequently appear on the E/S0 ridge line. Hence even without accreting new galaxies, $N_{200}$, the red-sequence richness, increases as more galaxies turn red. Various works have reported an increase of the number density of faint red-sequence cluster members toward low redshift \citep[e.g.][]{Loh08,Gilbank08,Rudnick09,Jaffe11,Vulcani11}. For example, \citet{Rudnick09} measure the optical cluster luminosity function of red-sequence galaxies at $z$$<$0.8. They find that at magnitudes brighter than $M^\star$, the luminosity function does not evolve much, suggesting that these cluster members are already in place. However, at fainter magnitudes, the luminosity function strongly increases towards lower redshift. This supports the view that cluster richnesses (defined with a $M^\star+1$ magnitude limit, as we do) may become larger towards lower redshift, in line with what our results suggest. There may be other processes that could cause an evolution of $N_{200}$. Mass segregation could lead to more bright cluster members within the inner one Mpc, which would boost $N_{\rm gal}$, leading to higher values for $r_{200}^{\rm gal}$ and $N_{200}$. Mergers of cluster members both brighter than $M^\star+1$ would lower $N_{200}$, but mergers of faint red-sequence members could increase it. The build-up of the red sequence, however, is likely the dominant effect. \\
\indent The redshift dependence of the mass-richness relation was also measured in \citet{Sheldon09I} for the maxBCG clusters, but due to the limited redshift range of that sample no change with redshift was found. However, in a study of the relation between X-ray luminosity and richness for the maxBCG clusters, \citet{Rykoff08} found that the X-ray luminosity at $z=0.28$ is twice as high as the X-ray luminosity at $z=0.14$. \citet{Becker07} studied the relation between velocity dispersion and richness for the same clusters, and found that the clusters at high redshifts systematically have higher velocity dispersions. Both \citet{Becker07} and \citet{Rykoff08} expected the main cause to be the evolution of the $N_{200}$ richness measure, implying a fractional decrease in $N_{200}$ of 30\%-40\% from $z=0.14$ to $z=0.28$. We ignored this effect when we compared our results in Fig. \ref{plot_n200m200}, but since the mean redshift of maxBCG clusters is 0.25, very close to the mean redshift of our low-redshift clusters, it is not important. If we ignore the potential evolution in halo mass, our results suggest a fractional decrease of $34\pm4$\% in $N_{200}$ from $z=0.27$ to $z=0.46$. \\
\indent \citet{Andreon14} measure the richness and mass for a sample of 23 very massive clusters with 0.15$<$$z$$<$0.55 within a fixed aperture of 0.5 Mpc. This `aperture' mass-richness relation does not evolve with redshift. Most of the accretion and quenching of new cluster members happens at the outskirts of galaxy clusters \citep[e.g.][]{Vanderburg15}, and may be missed when using an 0.5 Mpc aperture. Furthermore, the selection of red cluster galaxies and the computation of richness differ from our work, which may contribute to the apparent discrepancy between the results.


\section{Miscentring distribution  \label{sec_msc}}
\begin{figure}
  \resizebox{\hsize}{!}{\includegraphics{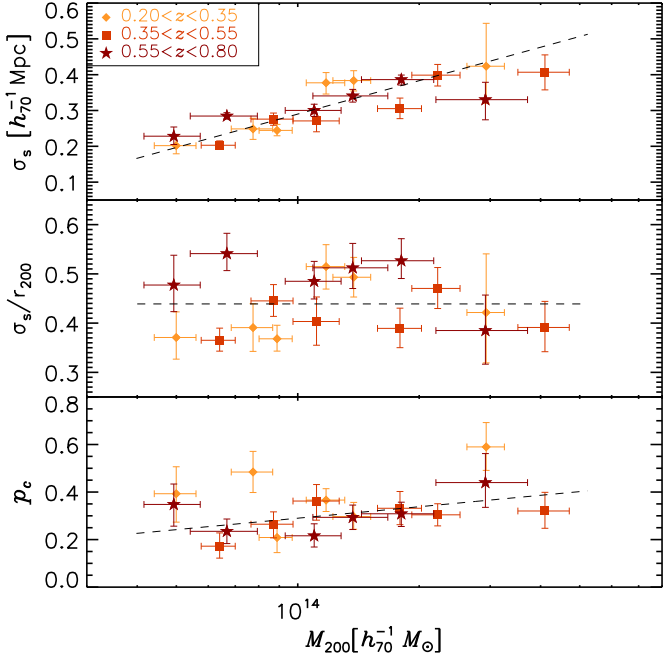}}
 \caption{Cluster mass versus the width of the miscentring distribution ({\it top}), versus the width divided by $r_{200}$ ({\it middle}) and versus the fraction of `correctly centred' BCGs, the ones that are located at the centre of the halo  ({\it bottom}). The black dashed lines show the fitted relation between miscentring parameter and mass, as described in the text.}
  \label{plot_miscent}
\end{figure}
By fitting the shear measurements together with the cluster-satellite correlation, we obtain, for the first time, tight constraints on the miscentring distribution of BCGs with respect to the centre of their dark matter haloes. Figure \ref{plot_miscent} shows that $\sim$30\% of our BCGs are located at the centre of the halo. The distribution of the miscentred BCGs is described by a 2D-Gaussian with a standard deviation that increases from $\sim$0.2 to $\sim$0.4 $h_{70}^{-1}$Mpc going from our poorest to our richest clusters; the ratio of width and $r_{200}$ is flat with mass. Since the miscentring parameters do not appear to evolve with redshift, we combine all our cluster samples and parametrise the relations as \mbox{$p_c=A_p+B_p \times (\log_{10}(M_{200})-14)$}, obtaining $A_p=0.29\pm0.02$ and \mbox{$B_p=0.16\pm0.07$}, and \mbox{$\sigma_s=A_\sigma+B_\sigma \times (\log_{10}(M_{200})-14)$}, finding $A_\sigma=0.29\pm0.01$ [$h_{70}^{-1}$Mpc] and \mbox{$B_\sigma=0.31\pm0.03$} [$h_{70}^{-1}$Mpc/$\log_{10}(h^{-1}_{70}M_\odot)$]. Since cluster members are spread over a larger volume in more massive clusters, we expect that the mass dependency is reduced when we consider the ratio $\sigma_{\rm s}/r_{200}$. Indeed we find this ratio is consistent with a constant, for which we obtain a value of $0.44\pm0.01$ for all samples combined. The average for the low-, intermediate- and high-redshift slices are $0.42\pm0.02$, $0.40\pm0.02$ and $0.51\pm0.02$, respectively. Note that we account for the errors on both the dependent and independent parameters in the fit. The high-redshift slice has an intrinsically broader miscentring distribution, which may reflect that more clusters are still undergoing mergers at that redshift, or that the BCG is more often misidentified due to increased photometric errors. This miscentring distribution is different from the one adopted in \citet{Johnston07}, which was based on mock catalogues determined from simulations that were fit to a range of observations. There, a much higher fraction of 60-80\% of the BCGs was found to be correctly centred. The width of the miscentring distribution was found to be 0.42 $h^{-1}$Mpc, larger than what we find. The lensing signal itself cannot discriminate between their miscentring distribution and ours, but the cluster-satellite correlation is able to break the degeneracy. \\
\indent It is possible that the difference between the miscentring distributions is the result of the different algorithms used to identify the BCG. This could be tested by applying both algorithms to the same set of simulations or data. However, part of the discrepancy could also be due to how the simulations used in \citet {Johnston07} are constructed. In these simulations, every dark matter halo is assumed to have a BCG at its centre. Hence misidentifying the BCG is the only reason why the centring fraction is less than 100\%. However, in reality, the central BCG may actually be displaced by a small amount from the centre of the halo. This would push the centring fraction down and lower the width of the miscentring distribution, more in line with our findings. Also, some BCGs may be star forming, leading to too blue colours to be selected as red-sequence member \citep{Bildfell08}. \\
\indent The location of BCGs in clusters has been studied in various other works. \citet{Skibba11} study mock catalogues and SDSS group catalogues and report that in 40\% of their groups with a mass larger than 5$\times$$10^{13}$ $h^{-1} M_\odot$, the brightest galaxy is a satellite galaxy instead of the central galaxy. Using N-body simulations, \citet{Martel14} find that the fraction where the brightest galaxy is not the nearest to the centre increases from $\sim$25\% to $\sim$50\%, with a higher miscentring fraction towards higher mass. \citet{Hoshino15} study the distribution of LRGs in the redMaPPer clusters \citep{Rykoff14} and find that 20-30\% of the brightest LRGs are not the central galaxy. As the central galaxy might be somewhat offset from the peak of the dark matter, these results might be in agreement with our results. Remarkably, both \citet{Skibba11} and \citet{Martel14} find that the centring fraction actually decreases with increasing mass, in contrast to what is assumed in \citet{Johnston07}. This is attributed in \citet{Martel14} to cluster mergers, which have occurred most recently in more massive clusters as they are the last ones to form. Our centring fractions do not show a clear trend with mass. \\
\indent \citet{George12} study the miscentring distribution in 129 X-ray-selected galaxy groups using their stacked weak-lensing signal, for different group centre definitions. In their model, all centres follow a 2D-Gaussian distribution, as their data does not require a correctly-centred component. Adopting the brightest group galaxy within $r_{200}$ as the group centre, they report $\sigma_s=24.8\pm12.0$ kpc, significantly smaller than the typical values for $\sigma_s$ that we find. Since the groups were X-ray selected, the sample may contain relatively more relaxed systems, whose BGGs could be closer to the centre of the dark matter distribution than for the full population of groups. \\
\indent \citet{Zitrin12} studied the miscentring distribution of BCGs in 10 000 SDSS clusters, under the assumption that light traces mass. Also in that work, a much narrower miscentring distribution is reported, with a typical width of 15 kpc. Since the BCG is usually by far the brightest galaxy in a cluster, this result may not be that surprising. \citet{Zitrin12} do not characterise the offset distribution of the misidentified clusters, which constitutes about 10\% of the sample. Part of the discrepancy between the results of \citet{George12,Zitrin12} and our findings may be caused by the fact that we include a correctly-centred component, as our data require it. Forcing all our BCGs to follow a 2D-Gaussian would lower $\sigma_s$. 


\section{Mass-concentration relation  \label{sec_mc}}
\begin{figure}
  \resizebox{\hsize}{!}{\includegraphics{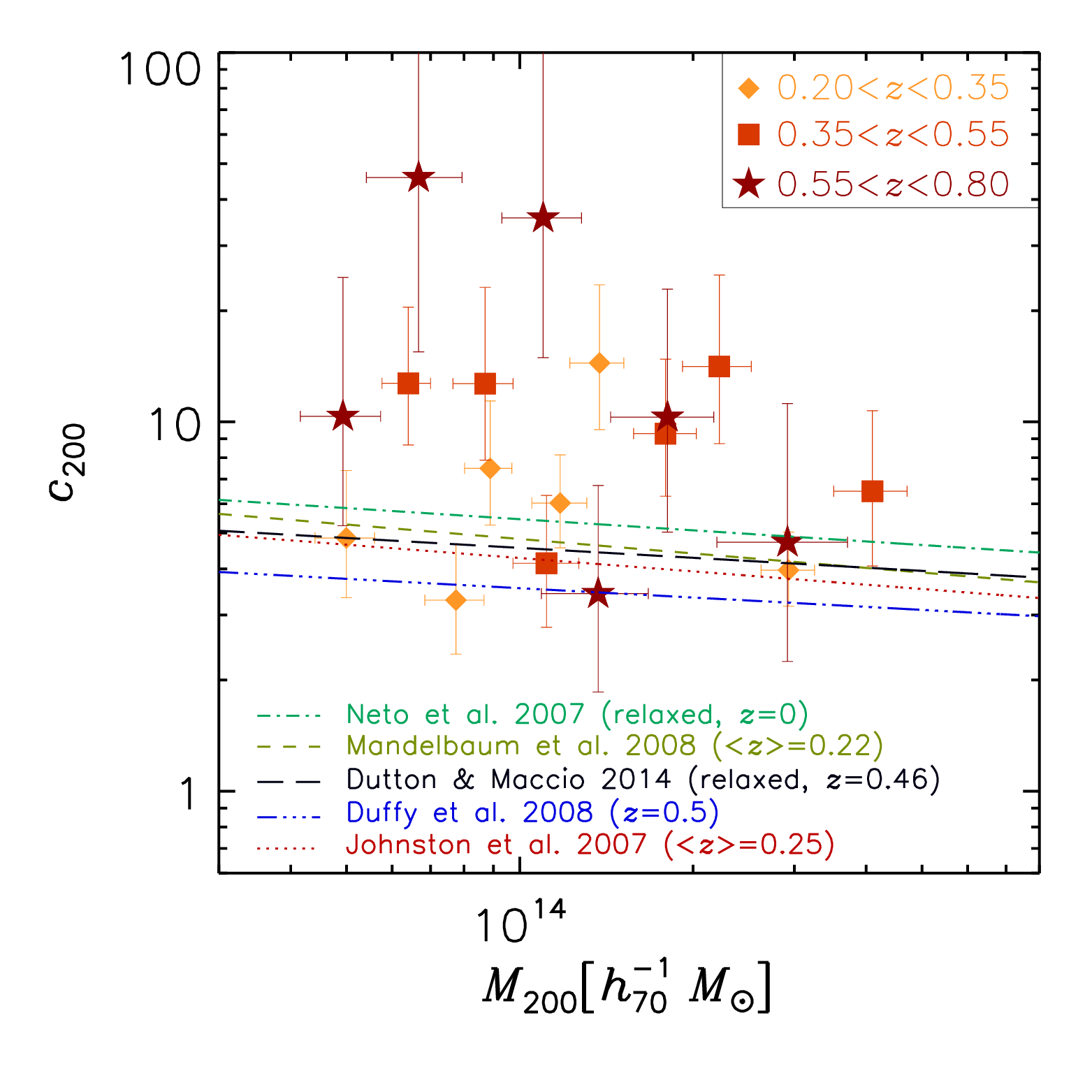}}
  \caption{Cluster mass versus concentration for the different redshift slices. The lines show the mass-concentration relations from \citet{Johnston07}, \citet{Mandelbaum08conc}, \citet{Neto07}, \citet{Duffy08} and \citet{DuttonMaccio14}, as indicated in the figure.}
  \label{plot_m200conc}
\end{figure}
Figure \ref{plot_m200conc} shows the relation between mass and concentration, together with a number of literature results: the mass-concentration relation of the maxBCG clusters from \citet{Johnston07}, at a mean redshift of 0.25; the results from \citet{Mandelbaum08conc}, who derived the mass-concentration relation by combining lensing measurements for L$*$-type galaxies, galaxy groups traced by LRGs and the maxBCG sample, for a mean redshift of  $z=0.22$; the relation of \citet{Neto07}, derived using the Millennium Simulation at $z=0$; the relation from \citet{Duffy08} at $z=0.46$, based on  large N-body simulations using the WMAP5 cosmology \citep{Komatsu09}; and, finally, the relation from \citet{DuttonMaccio14} at $z=0.46$, derived from N-body simulations using the Planck cosmology \citep{Planck14}. \\
\indent The concentrations for our low-redshift slice agree well with the literature results. For example, if we fit the mass-concentration relation from \citet{Johnston07} (\citet{Mandelbaum08conc}) to our results, we find a relative normalisation of $1.16\pm0.16$ ($1.03\pm0.14$), consistent with unity. The concentrations for our clusters in the 0.35$<$$z$$<$0.55 range are higher. If we fit the mass-concentration relation from \citet{DuttonMaccio14} (\citet{Duffy08}) at $z=0.46$, we find a relative normalisation of $1.78\pm0.31$ ($2.28\pm0.40$), 2-3$\sigma$ larger than unity. For the highest redshift slice, the relative normalisation is $1.67\pm0.54$ ($2.12\pm0.69$) with respect to the relation from \citet{DuttonMaccio14} (\citet{Duffy08}) at $z=0.67$. Note that we have ignored the error on the mass in deriving the relative normalisations, which should not matter much as the concentration changes only very weakly with mass. \\
\indent To investigate how sensitive our results are to the chosen prior, and to simultaneously fitting the cluster-satellite correlation, we look at how our results change when we only fit the lensing signal, using the priors from \citet{Johnston07}. The resulting relative normalisation of the mass-concentration relation of the 0.20$<$$z$$<$0.35 redshift slice becomes $0.85\pm0.08$ ($0.76\pm0.07$) with respect to the relation from \citet{Johnston07} (\citet{Mandelbaum08conc}). This is about 2$\sigma$ lower than our nominal result and therefore not significant. Figure \ref{plot_posterior2} shows that $q$ and $c_{200}$ are anti-correlated, such that forcing $q$ to high values (lowering the miscentring fraction) pushes $c_{200}$ down, in line with our results. Note that the constraints become tighter when we use the priors from \citet{Johnston07} because of the use of informative priors on $q$ and $\sigma_s$. Finally, we have also checked that our results are robust against moderate changes in modelling the two-halo term. Only if we apply a large boost to the 2-halo term, for example by using the non-linear power spectrum instead of the linear one, the best-fit concentrations become significantly higher. \\
\indent Our intermediate- and high-redshift clusters have a concentration that is higher than what is expected from dark-matter-only simulations. The question of whether clusters are over-concentrated with respect to simulations has received considerable attention in recent years \citep[e.g.][]{Umetsu11,King11,Oguri12,Auger13,Foex14,Umetsu14}. Most of these works focus on strong lensing systems, which are expected to have higher projected concentrations than typical clusters of similar mass and redshift \citep{Hennawi07,Oguri09,Oguri12}. On the other hand, weak-lensing studies of optically selected low-redshift galaxy clusters, as well as our 0.20$<$$z$$<$0.35 results, do not show a strong deviation from the mean mass-concentration relation predicted in $\Lambda$CDM cosmologies. \\
\indent Any red-sequence cluster finder that applies a spatial filtering, like ours, will have some preference for selecting over-concentrated clusters, that is, structures elongated along the line of sight. To check whether it is possible that our results are affected by selection effects, we estimate the completeness. If our sample is complete, we cannot systematically miss the under-concentrated clusters. We compute the cumulative halo mass function at the average redshift of each of the three redshift slices and multiply that with the volume in that slice. This gives us a crude estimate of the expected number of clusters above a given mass in the RCS2. Next, we use our mass-richness relation to assign each cluster a mass and compute the observed cumulative cluster mass function. Comparing this to the theoretical prediction gives us an estimate of the completeness. For our lowest redshift slice, our cluster sample is roughly complete at $M_{200}>10^{14}h_{70}^{-1}M_\odot$; for our intermediate-redshift clusters, we find a completeness of \mbox{70-80\%} in the same mass range; and for the high-redshift clusters, the completeness is less than 50\%. Note that these are crude estimates, but they show that it is possible that our intermediate- and high-redshift clusters are affected by selection effects. \\
\indent Another effect that might affect the concentrations are projections of pairs of clusters. Large-scale structure may be present between the clusters, which boosts the projected density at small scales. Additionally, the two parts of the projections will have lower masses and therefore on average higher concentrations, as the concentration is a decreasing function with mass, although this effect is small as the mass-concentration relation is fairly flat. If the projected separation of the two parts is small, both effects would increase the concentration. Larger projected separations are more likely, however, and would tend to lower the concentrations. Note that we find further evidence for the presence of selection and projection effects from the constraints on the halo bias, which we discuss in the following section. The bias on the concentration caused by triaxiality or halo substructure is expected to be much smaller \citep{Bahe11}. \\


\section{Satellite distribution  \label{sec_msat}}
\begin{figure}
  \resizebox{\hsize}{!}{\includegraphics{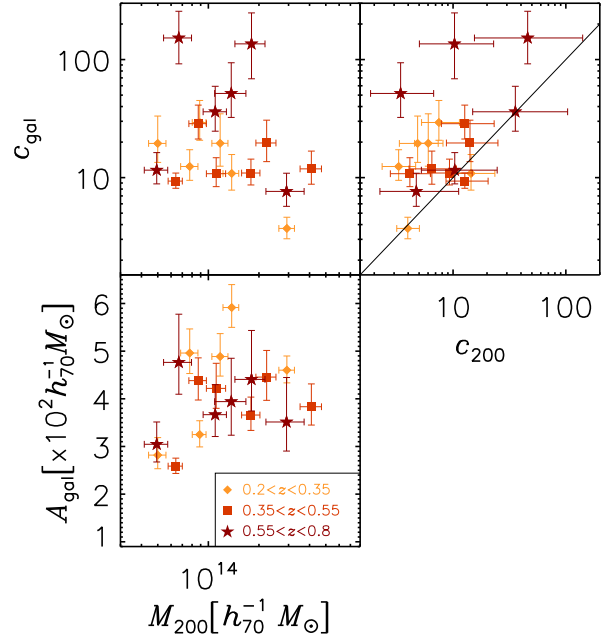}}
 \caption{Parameters describing the dark matter distribution (horizontal axis) versus parameters describing the distribution of satellites (vertical axis) for the different redshift samples. {\it (top left)}: halo mass versus concentration of the satellite distribution; {\it (top right)}: dark matter concentration versus concentration of the satellite distribution, with the one-to-one relation as solid black line; {\it (bottom left)}: halo mass versus scaling of the surface mass density to the number density of satellites. }
  \label{plot_satdist}
\end{figure}

We show the parameters that describe the satellite distribution in Fig. \ref{plot_satdist}. $c_{\rm gal}$ is the most sensitive parameter to how we model the two-halo term. Given the large uncertainties in modelling the one-to-two-halo transition regime, the results need to be interpreted with great care. The reason why $c_{\rm gal}$ is so sensitive, can be seen as follows. Different implementations of the two-halo term mainly change the signal at small scales, but not the large (linear) scales. When we for example increase the signal from the two-halo term at small scales, where there is overlap with the miscentred term, the one-halo profile needs to become steeper, which is done by forcing $c_{\rm gal}$ upwards. \\
\begin{table}
  \caption{Fit parameters of the relation between $A_{\rm gal}$ and $M_{200}$.}   
  \centering
  \begin{tabular}{c c c } 
  \hline
   & & \\
 Bin & $B$ & $\beta$  \\
   & [$10^{-2} h_{70} M^{-1}_{\odot}$] & [$\times10^2$] \\
   & & \\
  \hline\hline
   & & \\
0.20$<$$z$$<$0.35 & $4.1\pm0.2$ & $2.2\pm0.6$ \\
0.35$<$$z$$<$0.55 & $3.3\pm0.1$ & $2.1\pm0.5$ \\
0.55$<$$z$$<$0.80 & $3.8\pm0.3$ & $1.0\pm1.0$ \\
  \hline \\
  \end{tabular}
  \label{tab_agal}
\end{table} 
\indent Since a different implementation mainly results in a rescaling of $c_{\rm gal}$, we can only safely interpret the trends in the data, but not the values themselves. Our data suggests that the concentration of the satellite distribution decreases with mass. If we fit a linear relation between $\log_{10}(M_{200})$ and $c_{\rm gal}$, we obtain a non-zero slope of $-10.7\pm2.3$. Secondly, $c_{\rm gal}$ and $c_{200}$ are correlated. A correlation between $c_{200}$ and $c_{\rm gal}$ is expected, given that red-sequence galaxies trace the dark matter. Figure \ref{plot_posterior2} suggests that this is not the result of a degeneracy between the parameters. We quantify this correlation with the Pearson coefficient, which has a value of $0.67_{-0.14}^{+0.11}$. The error on the Pearson coefficient corresponds to the 68\% confidence intervals, obtained from \mbox{10 000} random realisations of the data, in which we draw new concentration values from a Gaussian whose mean equals the best-fit value and whose width corresponds to the observed error. \\
\indent Finally, our measurements show that $A_{\rm gal}$, the scaling between the model surface mass density and the number density of red-sequence members brighter than $M^\star+1$, increases with mass. Since the concentration of the satellite distribution is allowed to differ from that of the dark matter, this cannot be directly interpreted as a change in scaling between the surface mass density of dark matter and the projected density of red-sequence cluster members brighter than $M^\star+1$. $A_{\rm gal}$ merely serves as a nuisance parameter. For completeness, we parametrise this as \mbox{$A_{\rm gal}=B+ \beta\times \log_{10}(M_{200}/10^{14})$} and list the best-fit values in Table \ref{tab_agal}. These results are not sensitive to moderate changes in the two-halo term. 


\section{Mass-bias relation  \label{sec_mb}}
\begin{figure}
  \resizebox{\hsize}{!}{\includegraphics{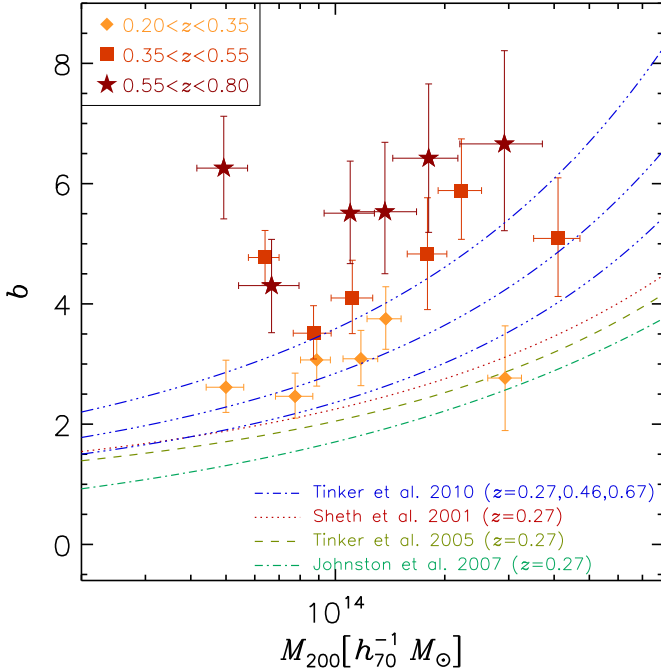}}
  \caption{Cluster mass versus bias from our halo model fits to the lensing and cluster-satellite signals for our cluster samples.  The lines show the results from \citet{Tinker10}, \citet{Sheth01}, \citet{Tinker05} and \citet{Johnston07}, as indicated in the figure.}
  \label{plot_m200bias}
\end{figure}
We show the relation between cluster mass and bias in Fig. \ref{plot_m200bias}, together with a number of relations from the literature, including the relation derived in \citet{Johnston07} for the maxBCG clusters, the prediction from \citet{Sheth01} based on an ellipsoidal collapse model and calibrated on numerical simulations, with further refinements presented in \citet{Tinker05}, and a new functional form for the mass-bias relation from \citet{Tinker10}, derived from a large set of collisionless numerical simulations based on $\Lambda$CDM. For clarity, we only show the relations at a redshift of $z=0.27$, except for the most recent one, that is, the \citet{Tinker10} relation; the relation with the lowest (highest) amplitude is for $z=0.27$ ($z=0.67$). \\
\indent Our bias values increase with redshift as expected \citep[for a compilation of redshift-dependent bias models, see Fig. 1 of][]{Clerkin15}, but they are somewhat larger than the relations from the literature. Since our clusters do not span a wide range of mass, we only constrain the relative normalisations with respect to these relations. Fitting the relation of \citet{Tinker10} (\citet{Johnston07}) at $z=0.27$ to our low-redshift results, we find a relative normalisation of 1.24$\pm$0.08 (1.73$\pm$0.11). For the intermediate- and high-redshift slices, we find a relative normalisation of 1.45$\pm$0.08 and 1.51$\pm$0.11 with respect to the relation of \citet{Tinker10} at $z=0.46$ and $z=0.67$, respectively. Note that we have again ignored the errors on the mass in deriving these relative normalisations, which should not matter much as the bias is only a weak function of mass. \\
\indent If we use the priors from \citet{Johnston07} and only fit the lensing signal, the errors on the bias blow up since the lensing signal at large scales is noisy. \citet{Johnston07} fit the lensing signal out to 30 $h^{-1}$ Mpc, which provides more constraining power on the amplitude of the two-halo term. Since the large-scale lensing signal is generally weak and therefore more susceptible to biases due to potentially remaining systematics in the shape catalogues, we deem that further testing of the robustness of our lensing signal is required before we can extend the fitting range. Also the cluster-satellite correlation becomes harder to measure reliably at larger scales, as it becomes increasingly sensitive to the background density. \\
\indent We have tested how our results change when we implement a different prescription for the two-halo term. The biases are fairly robust against moderate changes. Only if we apply a large boost to the two-halo term, for example by replacing the linear power spectrum with the non-linear one (as is done in \citep[][]{Cacciato09,VanUitert11,Leauthaud12}), the values become $\sim$10\% lower. Next to that, we note that our halo models on average underpredict the cluster-satellite correlation at \mbox{1-2 $h_{70}^{-1}$Mpc}, which may bias the halo bias high. \\
\indent The difference between our bias values and the literature relations is larger for our intermediate- and high-redshift clusters. This may be caused by the same combination of effects that might explain the high concentrations. If our cluster finder preferentially selects overdense clusters, the large-scale structure will also be preferentially oriented along the line-of-sight. Similarly, if our cluster is actually a projection, more additional structure may be present along the line-of-sight. Both would lead to a larger projected clustering of matter, and hence a larger bias.


\section{Conclusion \label{sec_concl}}
\hspace{4mm} We present the results from a combined weak-lensing and cluster-satellite correlation analysis of $\sim$$10^{4}$ clusters in the RCS2. These clusters span a range of 0.2$<$$z$$<$0.8 in redshift and have typical masses $M_{200}$$>$$5\times10^{13} h_{70}^{-1} M_{\odot}$. We divide the clusters in three redshift slices and six richness bins, and measure the average lensing signal and cluster-satellite correlation for each sample. Satellites are identified as all red-sequence galaxies at the cluster redshift brighter than $M^\star+1$. We model the signals simultaneously using a cluster halo model, in which we account for the miscentring distribution and the scatter between richness and mass. From these fits, we obtain the masses, the concentrations of the dark matter and of the satellite distribution, the bias, and the miscentring parameters. \\
\indent We parametrise the relation between mass and richness as \mbox{$M_{200}=A\times(N_{200}^{\rm{corr}}/20)^{\alpha}$} and find \mbox{$A=(15.0\pm0.8) \times10^{13} h_{70}^{-1} M_{\odot}$} and $\alpha=0.73\pm0.07$ for our low-z clusters at 0.2$<$$z$$<$0.35. At intermediate redshift (0.35$<$$z$$<$0.55), we find a higher normalisation of \mbox{$A=(22.5\pm1.7) \times10^{13} h_{70}^{-1} M_{\odot}$}. Passive evolution and halo mass growth would lead to a higher normalisation at lower redshift, opposite to what we find. Hence we expect that this trend is driven from a fractional increase of $N_{200}$ towards lower redshift, caused by the build-up of the red sequence. Similar trends were observed in the analyses of maxBCG clusters \citep{Becker07,Rykoff08}. \\
\indent Our measurements provide tight constraints on the cluster miscentring distribution. Only $\sim$30\% of our BCGs are located at the centre of the halo; the remaining BCGs follows a 2D-Gaussian, whose width is approximately \mbox{0.2 $h_{70}^{-1}$Mpc} at $5\times10^{13} h_{70}^{-1} M_{\odot}$ and increases to \mbox{0.4 $h_{70}^{-1}$Mpc} for our most massive systems. The ratio of the width and $r_{200}$ is flat with mass, with an average value of $0.44\pm0.01$. Our miscentring fraction is higher than what is typically reported in the literature, which might be caused by the commonly made assumption that the central galaxy always resides exactly at the centre of the dark matter distribution, and any miscentring is caused by not correctly identifying the central galaxy. In reality, even the central galaxy may be offset from the centre of the halo, as allowed for in our modelling, leading to smaller centring fractions. \\
\indent The mass-concentration relation of our low-z cluster sample agrees well with the predictions from numerical simulations, but for our intermediate- and high-redshift clusters, a higher normalisation is preferred. We hypothesise that this is the result of two effects: a selection effect, as our cluster finder preferably selects overdense systems; and a projection effect, where two clusters are located close to the line of sight and counted as one. We find further evidence for these effects from the constraints on the bias, which show a similar trend: fair agreement with numerical simulations at low redshift, but a preference for higher values at higher redshifts. \\
\indent The concentration of the satellite distribution decreases towards higher mass. It is correlated with the concentration of the dark matter. The corresponding Pearson coefficient has a value of $0.67_{-0.14}^{+0.11}$. \\
\indent We have tested the robustness of our results against various changes in our halo model, such as including different implementations of the two-halo terms and the use of different priors in the fit. These tests show that all but one parameter are robustly extracted. Only $c_{\rm gal}$ changes considerably when the two-halo term is modelled differently, and hence their values need to be interpreted with care. Since a different implementation mainly leads to a rescaling, the decrease of $c_{\rm gal}$ with mass and the correlation between $c_{\rm gal}$ and $c_{200}$ are a robust find of this work. \\
\indent The next step is to repeat our analysis on mock data. This is needed for a better characterisation of our cluster sample, in terms of cluster completeness, purity, and the assignment of the cluster's centres. Without that, a robust cosmological exploitation of our measurements via the halo mass function is not possible. In addition, mocks will also be very useful to help interpret our current findings, for example whether selection and projection effects can explain the high concentration and bias values that we find, or whether there is another cause. Finally, we plan to use our cluster sample with its red-sequence members for a variety of follow-up projects, including a range of alignment studies.


\paragraph{Acknowledgements \\ \\} 
\indent We would like to thank Peter Schneider and Erica Ellingson for their comments on an earlier version of this work. EvU acknowledges support from a grant from the German Space Agency DLR, from a Marie Curie International Reintegration Grant and from an STFC Ernest Rutherford Research Grant, grant reference ST/L00285X/1.  DG acknowledges support from the National Research Foundation of South Africa. HH acknowledges support from a Marie Curie integration grant and a VIDI grant from the Nederlandse Organisatie voor Wetenschappelijk Onderzoek (NWO). The RCS2 project is supported in part by grants to HKCY from the Canada Research Chairs program and the Natural Science and Engineering Research Council of Canada. \\
\indent This work is based on observations obtained with MegaPrime/MegaCam, a joint project of CFHT and CEA/DAPNIA, at the Canada-France-Hawaii Telescope (CFHT) which is operated by the National Research Council (NRC) of Canada, the Institute National des Sciences de l'Univers of the Centre National de la Recherche Scientifique of France, and the University of Hawaii. We used the facilities of the Canadian Astronomy Data Centre operated by the NRC with the support of the Canadian Space Agency. 

\bibliographystyle{aa}

\begin{appendix}


\section{BCG selection \label{ap_bcg}}
\begin{figure*}
  \resizebox{0.94\hsize}{!}{\includegraphics{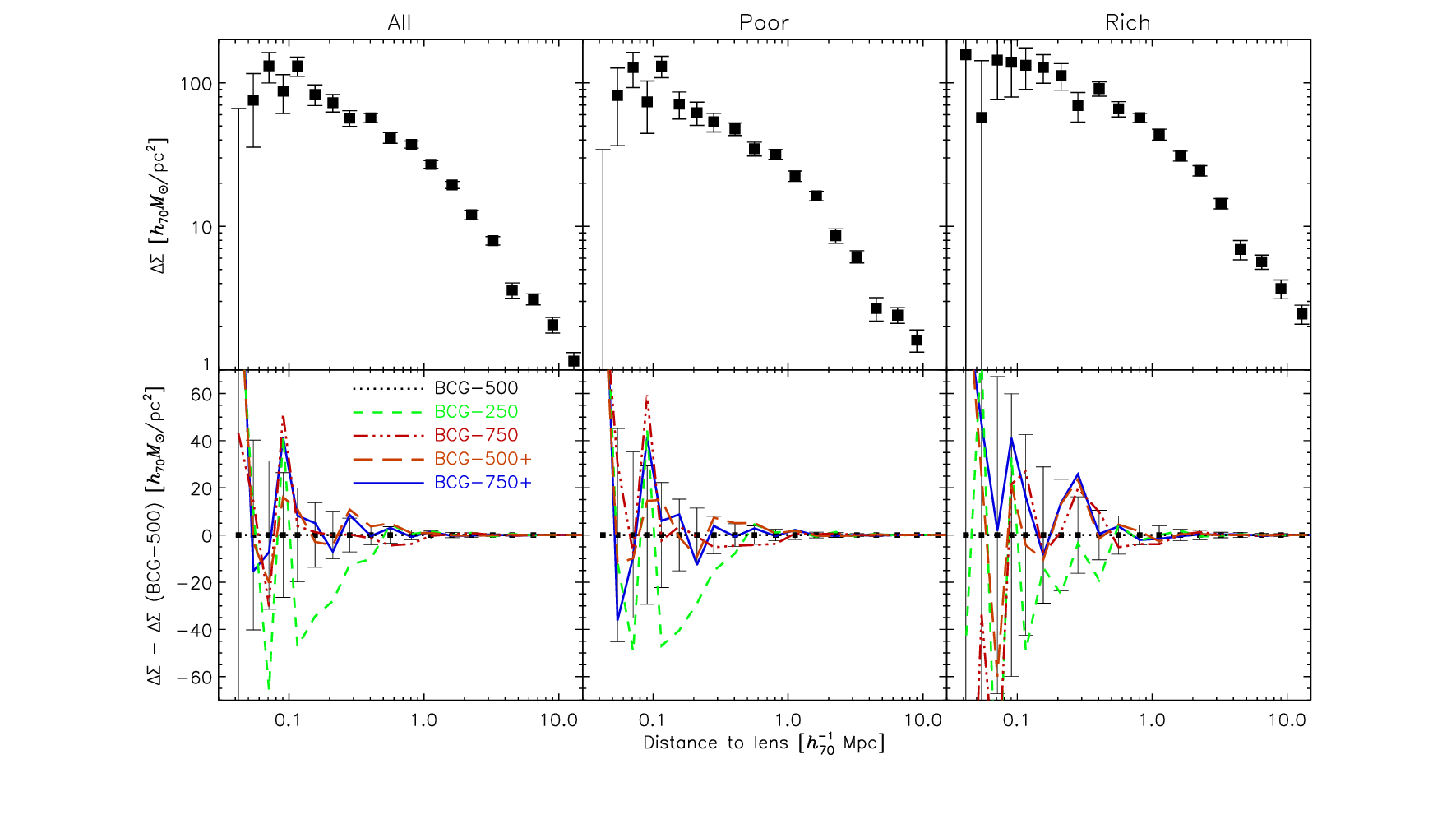}}
  \caption{Shear signal around BCG candidates selected with the reference `BCG-500' algorithm (top panels). The bottom panels show the difference in lensing signal between various BCG selection algorithms and the reference. The left-hand column shows the results for all clusters, the middle one for poor clusters and the right-hand panel for rich clusters. The error on the difference is simply approximated by the error on the reference shear measurement - the true error will only be slightly larger due to the large covariance of the results. The lensing signal around the `BCG-250' is clearly smaller at small projected separations, indicating that a higher fraction of these BCGs are miscentred.}
  \label{plot_bcg_shear}
\end{figure*}
\begin{figure*}
  \resizebox{0.94\hsize}{!}{\includegraphics{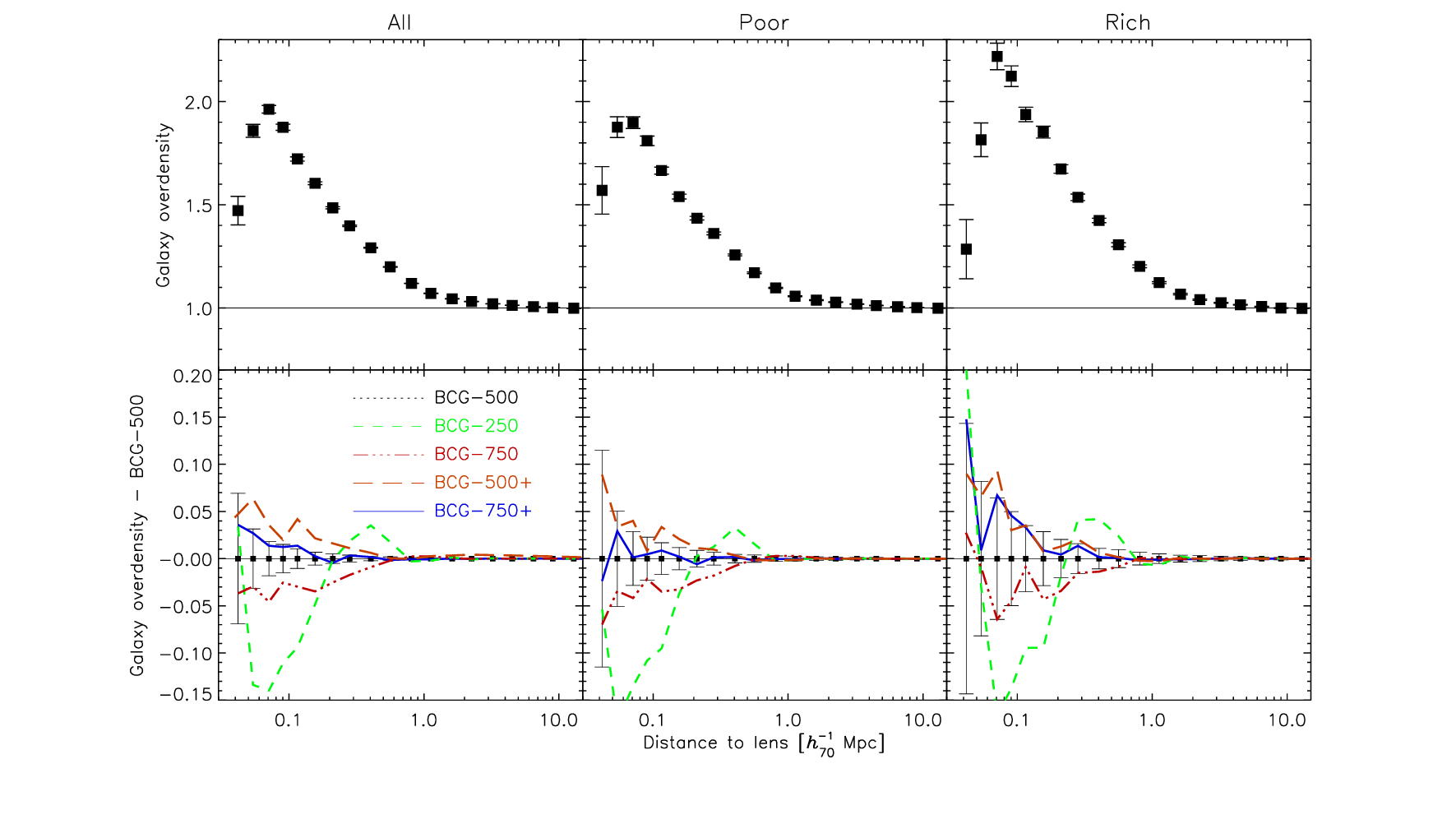}}
  \caption{Fractional galaxy overdensity around BCG candidates selected with the reference `BCG-500' algorithm (top panels). The bottom panels show the difference in galaxy overdensity between various BCG selection algorithms and the reference. The left-hand column shows the results for all clusters, the middle one for poor clusters only and the right-hand panel for rich clusters only. The error on the difference is simply approximated by the error on the reference overdensity measurement - the true error will only be slightly larger due to the large covariance of the results. The `BCG-500+' leads to the highest, most concentrated peak, indicating that these BCGs are closest to the centre of the cluster member distribution.}
  \label{plot_bcg_overd}
\end{figure*}
Both the stacked shear signal of all clusters and the total galaxy overdensity signal around the centre given by the cluster finder, reveal that the clusters have a broad miscentring distribution. In principle, if our cluster halo model is sufficiently flexible, our results should not critically depend on the choice of the cluster centre. However, given that the cluster miscentring distribution is somewhat uncertain, it is better to try to optimise the definition of the centre. \\
\indent To improve the centring of the clusters we attempt to identify the BCG. For this purpose, we use the catalogue of red-sequence galaxies that were used to identify the cluster. We try a number of different schemes for identifying the BCG and compare the stacked shear signal and galaxy overdensity signal. The most successful prescription will result in a maximal shear and galaxy overdensity signal at small scales; miscentring basically shifts power to larger projected separations.\\
\indent We start with identifying the brightest red-sequence candidate in the $z$-band at the cluster redshift within 250 $h_{70}^{-1}$ kpc, 500 $h_{70}^{-1}$ kpc and 750 $h_{70}^{-1}$ kpc from the original cluster centre and adopt it as the candidate BCG. The total lensing signal and galaxy overdensity around these BCG selections are shown in Fig. \ref{plot_bcg_shear} and \ref{plot_bcg_overd},  and are labelled `BCG-250', `BCG-500' and `BCG-750', respectively. The shear signal around the `BCG-250' BCGs is clearly lower at small scales than for the other options. Due to the large errors in the shear signal at small scales, there is no obvious difference in the shear signals between the `BCG-500' and `BCG-750' selections. However, the galaxy overdensity shows that the `BCG-500' BCGs better coincides with the peak of the galaxy distribution, and hence are better centred. \\
\indent A visual inspection of a number of rich clusters reveals that in some cases, the brightest galaxy is quite far offset from the distribution of cluster members. Hence to improve the `BCG-500' and `BCG-750' selections, we compute the weighted $z$-band luminosity centre of the cluster using all red-sequence candidates that are located within 500 $h_{70}^{-1}$ kpc from the original cluster's centre. If there exists a second-bright red-sequence candidate member at the cluster redshift that is at most 0.5 magnitude fainter than the first selected BCG, but that is located at least 100 $h_{70}^{-1}$ kpc closer to the centre of light, we adopt it as the BCG and consequently as the new centre of the cluster. This occurs in 30\% and 34\% of the `BCG-500' and `BCG-750' selections, respectively. The resulting BCG selections are called `BCG-500+' and `BCG-750+'. From Fig. \ref{plot_bcg_overd} we observe that the resulting BCG catalogues are indeed better centred. The one with the highest galaxy overdensity is `BCG-500+'. We therefore adopt the BCG candidates from this algorithm as the centres of the clusters. The richness estimates of the clusters are recomputed using the new centres. \\


\section{Impact of cluster redshift scatter \label{ap_zscat}}
\begin{figure}
  \resizebox{\hsize}{!}{\includegraphics[angle=270]{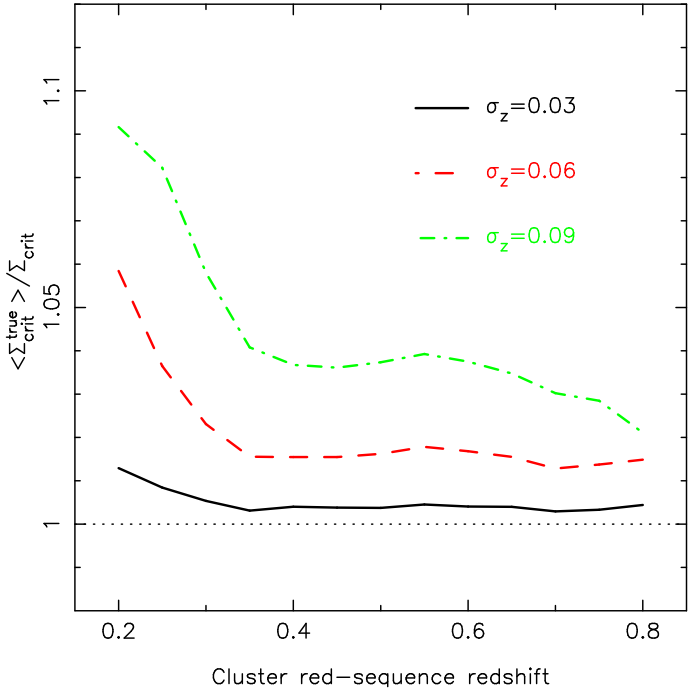}}
  \caption{Impact of redshift scatter on the lensing efficiencies.}
  \label{plot_scatSC}
\end{figure}

\begin{figure*}
  \resizebox{\hsize}{!}{\includegraphics[angle=270]{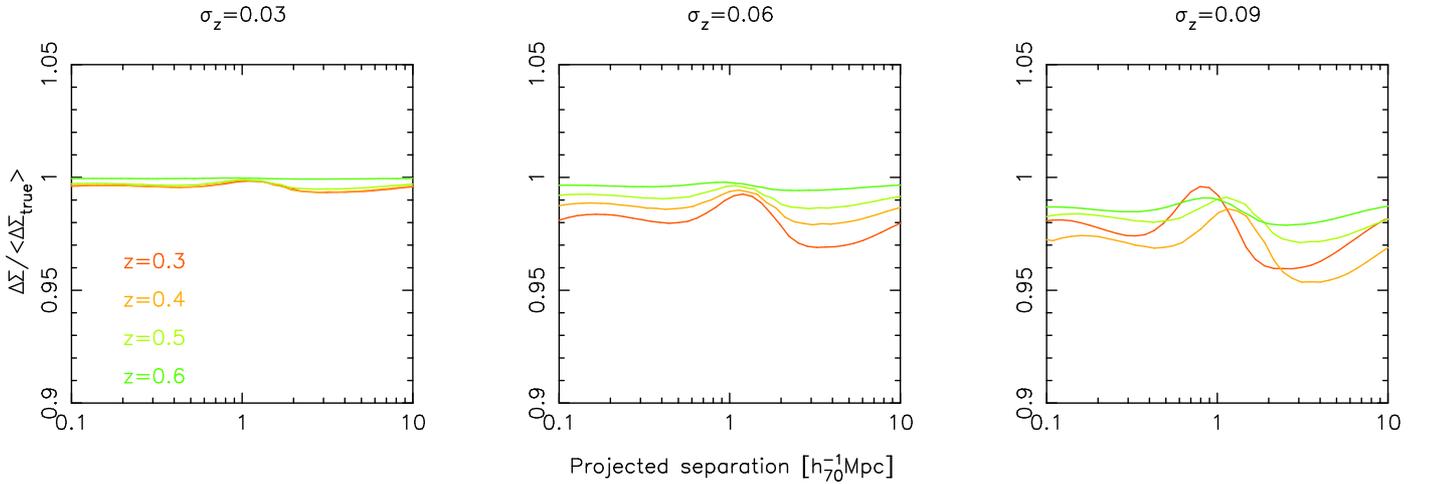}}
  \caption{Impact of redshift scatter on the radial profile of the model signal.}
  \label{plot_scatRadAv}
\end{figure*}

The assigned `red-sequence' redshifts of the clusters have a certain scatter with respect to their actual redshifts. This is caused by intrinsic scatter in the red sequence, uncertainties in the background subtraction, contamination of fore- and/or background galaxies, and noise. Redshift scatter affects the lensing analysis in three ways: it biases the adopted lensing efficiencies, it causes a radial smoothing of the lensing signal, and, if the signal is non-linearly redshift dependent, also an additional smoothing of the signal in that direction.\\
\indent The lensing efficiency does not linearly scale with lens redshift. Hence if the actual redshifts are scattered compared to the adopted ones, the average over these true lensing efficiencies does not equal the average over the lensing efficiencies of the adopted redshifts. We simulate the impact as follows. For a given cluster `red-sequence' redshift, we assume that the true redshift distribution follows a Gaussian with a certain width (this is identical to assuming that the true distribution of redshifts is flat, and the `red-sequence' redshifts follow a Gaussian distribution around the true value). We draw a large number of true cluster redshifts from this Gaussian, and for each we compute $\Delta\Sigma^{\rm true}_{\rm crit}$. If the redshift that is drawn is lower than 0.05 or higher than 1, we disregard it. We average these critical surface mass densities, and compare it to the $\Delta\Sigma_{\rm crit}$ of the input `red-sequence' redshift. We plot the ratio in Fig. \ref{plot_scatSC}.\\
\indent The average of the distribution of true $\Delta\Sigma_{\rm crit}$ is higher than the value of the single `red-sequence' redshift (the one we would use). The size of the bias increases at lower redshifts and for higher values of the scatter. For our cluster sample, the estimated size of redshift scatter is $\sim$0.03. The difference between the mean of the true critical surface mass densities and the one we used is smaller than one per cent over the range of cluster redshifts considered. This is considerably smaller than the statistical error, and we therefore ignore the effect. \\
\indent Next we investigate the impact of the radial smoothing due to redshift scatter. We only focus on the lensing signal here; the cluster-satellite clustering should be affected in a similar way. We create a typical cluster lensing profile with our halo model for a certain `red-sequence' redshift at fixed physical separations. Next we assume that the true redshift distribution follows a Gaussian with a certain width and draw true redshifts from that. For each redshift, we compute the signal at the same angular separation (but corresponding to different physical separations), and do the radial binning assuming the redshift is the adopted `red-sequence' redshift. Hence we radially smooth the model profile. Then we compute the ratio of the average of this smoothed shear profile and the input profile, and show it in Fig. \ref{plot_scatRadAv}.\\
\indent In general, the smoothed profile is larger than the profile at a fixed `red-sequence' redshift. The difference is largest at low redshifts, and increases when the redshift scatter increases. However, the actual ratio obviously depends on the model profile that we have assumed. Deriving a correction factor in a model independent way is therefore not obvious, and we refrain from doing so. It is also not necessary, given that the impact on the signal is of order a per cent, smaller than our statistical errors. \\
\indent Finally, there is the effect of additional smoothing in the redshift direction. Given that the scatter is much smaller than the redshift bin size used in this work, this effect is negligible. 

\end{appendix}

\end{document}